\documentclass[preprint]{aastex62}
\begin{document}

\author{E. Jourdain}
\affil{ CNRS; IRAP; 9 Av. colonel Roche, BP 44346, F-31028 Toulouse cedex 4, France\\
 Universit\'e de Toulouse; UPS-OMP; IRAP;  Toulouse, France\\}
\author{J.P. Roques}
\affil{ CNRS; IRAP; 9 Av. colonel Roche, BP 44346, F-31028 Toulouse cedex 4, France\\
 Universit\'e de Toulouse; UPS-OMP; IRAP;  Toulouse, France\\}
\title{2003-2019 Monitoring of the Crab  emission through  \textit{INTEGRAL} SPI, or vice versa.}

\begin{abstract}
The Crab Nebula is used by many instruments as a calibration source, in particular at high energy, where it is one of the brightest celestial object.
The spectrometer \textit{INTEGRAL} SPI (20 keV - 8 MeV), in operation since October 2002, offers a large dataset dedicated to this source, with regular campaigns planned twice per year.
We have analyzed the available data to quantify the source behavior on a long term scale and examine the stability level on  timescales from hour to years. As a result,  the source flux variability appears to be contained within less than $\pm$5\% around a $\sim$ 20 yr mean value, for broad bands covering the 20 keV - 400 keV energy domain, above which statistic limits any firm conclusion. In term of spectral shape, the Band model provides a good description of the observed emission between 20 keV and 2.2 MeV. The averaged spectrum best fit parameters correspond to a low energy slope of 1.99 $\pm$ 0.01, a high energy slope of -2.32 $\pm$ 0.02 and a characteristic energy $E_c$ of 531 $\pm$ 50 keV to describe the curvature joining both power laws. The spectral parameters have then been determined on the revolution timescale ($\sim$ 1 to 2 days)  and their  steadiness confirms the source emission stability.

As a complementary result, this study demonstrates that the SPI instrument efficiency  remains within 5\% of its initial value, after  17 years of operation.
\end{abstract}

\section{Introduction}
The Crab supernova remnant  is  one of the rare persistent sources  bright enough to provide significant signal to noise ratio in the Hard X-ray domain ($\sim$ 100 keV region) on the hour timescale. In addition, its emission appears  steady along the time, with minimal variations on a few year timescale ($\sim$ 2-3\% per yr), around a mean flux which remains stable over longer timescales \citep{crabvar}. Flux variabilities, including flares, are observed at higher energy (100 MeV and above, \citet{flareLAT}) without any impact at lower energy.
Also, the nebula extends over $\sim$ 5 arcmin, and can be considered as a punctual source for the coded mask instruments operating above 30 keV (i.e. all past and current hard X-ray instruments, except NuSTAR \citep{CrabNustar2015}). 
 All these properties make it the best tool to test, validate or correct instrument's  performance after launch.\\
Conversely, the assessment of the stability of the Crab  emission relies on the (cross-)calibration and stability of the measuring instruments.   
In this paper, we aim to disentangle both effects (source variability and
instrument evolution) using the observations gathered above 20 keV, by the SPI spectrometer aboard the \textit{INTEGRAL} observatory. This is possible because, unlike many instruments, SPI is not calibrated against an ad hoc Crab index: its response matrices are based on extensive ground calibrations and Monte-Carlo simulations. 
Since its launch at the end of 2002, \textit{INTEGRAL}  has regularly pointed toward the Crab Nebula (at least every 6 months) to check the instruments' performance. This provides an exceptional dataset to fulfill this two-fold objective.

In the following sections, we give a brief description of the instrument and dataset, then detail the photon selection criteria (including multiple events) and the analysis procedure. The results obtained are then presented (section \ref{resul}), with a discussion on the  evolution of the instrument efficiency and source variability (in flux and shape), respectively. Additional informations are given on the potential emission of a 511 keV annihilation line (section \ref{511}) and the instrument efficiency around 1.5 MeV (section \ref{40K}). Finally, we conclude with 
a few words about the  SPI instrument performance and the standard candle status of the Crab Nebula, in view of future cross-calibration efforts.

\section{Instrument, Observations and Data Analysis}
\subsection{The 'Spectrometre Pour Integral'}
 The  SPI instrument \citep{Vedrenne03} operates from 20 keV up to a few MeV, with a coded mask aperture associated to a detector plan consisting in 19 High purity Gemanium crystals. The various calibration runs conducted to characterize the instrumental response  are reported  in \citet{attie03}. The response matrices have been built through Monte-Carlo simulations, validated against ground measurements \citep{sturner03}. An overview of the in-flight performance can be found in \citet{roques03}, while a comprehensive description of the data analysis procedure can be found in \citet{crab09} with an update in \citet{roques19}. We just remind that the \textit{INTEGRAL} observations are organized  by revolutions (or orbits) which last $\sim$ 3 days, including 0.5 day of interruption due to radiation belts.  Each revolution consists in a succession of science windows (scw), lasting $\sim$ 2-3 ks.  The pointing direction changes from one scw to the next, by steps of $\sim 2^{\circ}$, following  pre-defined patterns. The complete information is available on the ESA INTEGRAL website (https://www.cosmos.esa.int/web/integral/schedule-information). Least, let's mention that specific annealing procedures are conducted more or less every six months to recover the energy resolution \citep{roques03}.

\subsection{The Data set}
 The following  analysis is based on the regular calibration campaigns performed twice a year, from the launch up to 2019 ones. Note that all these observations are made public immediately. By reason of specific calibration purpose, most of the Crab  observations before 2013 are performed with peculiar patterns, in order to investigate the transparency of the IBIS mask corners, and the response matrices for large off-angles for both SPI and IBIS instruments. These observations  have been used to demonstrate that data collected from sources up to $12^{\circ}$ from the INTEGRAL pointing axis (i.e. coded fraction $\sim$ 50\%) give reliable results.
 
 A careful check of both background evolution and goodness of the statistical criteria along the analysis process results in removing exposures affected by radiation belt exit or entry, loss of data, solar activity or other very bright transient events. For instance, all the observations  including bright outbursts of the nearby pulsar A0535+26 have been ignored in this work. In addition, revolutions encompassing less than 20 'good' scws have been excluded, since the determination of the background (and thus source) flux may be less accurate.
The final dataset thereby encompasses 56 revolutions corresponding to $\sim 3600$ science-windows (scw), and a total of 7 Ms. 
To ease the data analysis, the total dataset has been split into two sub-datasets, before and after revolution 1170 resp. (see below).

\subsection{Main features of the data analysis}\label{DataAn}

The flux extraction relies on a model fitting procedure, including both source and background components. For the Crab observations, the sky model is particularly simple: the Crab Nebula and the Be pulsar A0535+26. Except during flaring episods, A0535+26 is much fainter that the Crab (by a factor $>$ 50), and its variability  around its mean flux is limited. Consequently, the periods with strong outbursts being excluded, the A0535+26 flux can be  considered as constant on the revolution timescale during our analysis.  Concerning the Crab, the flux extraction has been performed on the scw and revolution timescales, with a view to achieving different scientific objectives. \\
The peculiar issue of the electronic noise and associated pile-up in the high energy part has been taken into account  following the recommendations described in \cite{roques19}. For the first part of the dataset (revolutions before 2012 May 13th = MJD 56060 = rev 1170), events without the PSD\footnote{A Pulse Shape Discrimination system (PSD) was implemented, in addition of the  main electronic chain, to disentangle  single and multi-site interactions and reject the formers, which are produced by local radioactivity. After launch, it appears that the rejection method results in a decrease of the SPI sensitivity and has thus been disabled. However, the PSD electronic chain remains operational and provides a specific flag to each analyzed photon. See \cite{roques03} for more details.}  flag are  ignored above 650 keV, and the fluxes are corrected from the PSD electronic chain efficiency ($\sim$ 85\%)\footnote{Tests on the PSD electronic chain have been performed from revolution  420 (2006-03-22, MJD 53816) to revolution 445 (2006-06-07, MJD 53893), implying a PSD threshold  fixed to 750 keV and a PSD dead time increasing to 25\% (see Table 1 in \cite{roques19}). This specific configuration has been taken into account for the only concerned revolution, namely revolution 422, around MJD 53824.}. Below 650 keV, we apply the pile-up correction factor, estimated to be $\sim$ 0.14 \%  of the source flux measured at 60 keV. This correction becomes negligible below 200 keV.
For the second part of the mission lifetime (after 2012 May 13th), the PSD configuration corresponds to a low energy threshold value of 400 keV. We thus select  PSD flagged events above this energy, and apply the appropriate corrections for PSD efficiency or pile-up, for fluxes above and below 400 keV, respectively.\\
An additional information is contained in the Multiple Events (ME). These events correspond to photons which deposit energy in more than one detector and are used for  polarimetric studies. In practice, we consider only double events, hereafter called ME2, produced by  high energy photons which undergo a Compton interaction in one detector and escape toward a neighbouring detector where they are photo-absorbed. 
The ME2 contribution is negligible below 90 keV and less than 20 $\%$ of the total flux at higher energy. However, it will be used in this paper to ascertain the high energy results.
The  flux extraction for ME is performed in a similar way than for the 'one-detector' events  while  the spectral deconvolution is completed with the appropriate matrices \citep{sturner03}.\\

To monitor the temporal evolution of the Crab Nebula over the INTEGRAL mission lifetime, fluxes have been extracted in four broad bands between 24 and 650 keV, to produce light curves on the revolution and scw timescales. Then, an analysis  in 41 narrower energy bins, covering the 24-2200 keV domain, has been performed on the revolution timescale, for the spectral studies.  
Fluxes from Multiples Events (ME2) have  been extracted between 90 and 2200 keV (16 channels), to be fit simultaneously with the corresponding one-detector event spectra. However, the ME efficiency suffering large uncertainties up to 170 keV, spectral fits will exclude the first five channels.  
At the end, we get a total of 56 individual spectra in the two datasets. 

For practical reasons (different correction parameters), the two observational periods  before and after the rev 1170 (22 and 34 revolutions resp.) have been analyzed separately.\\

\section{results}\label{resul}
\subsection{Broad band analysis}
\begin{figure}[ht]
\includegraphics[angle=0,width=9cm,height=22cm,trim= 0.cm 0cm 0cm 2cm]{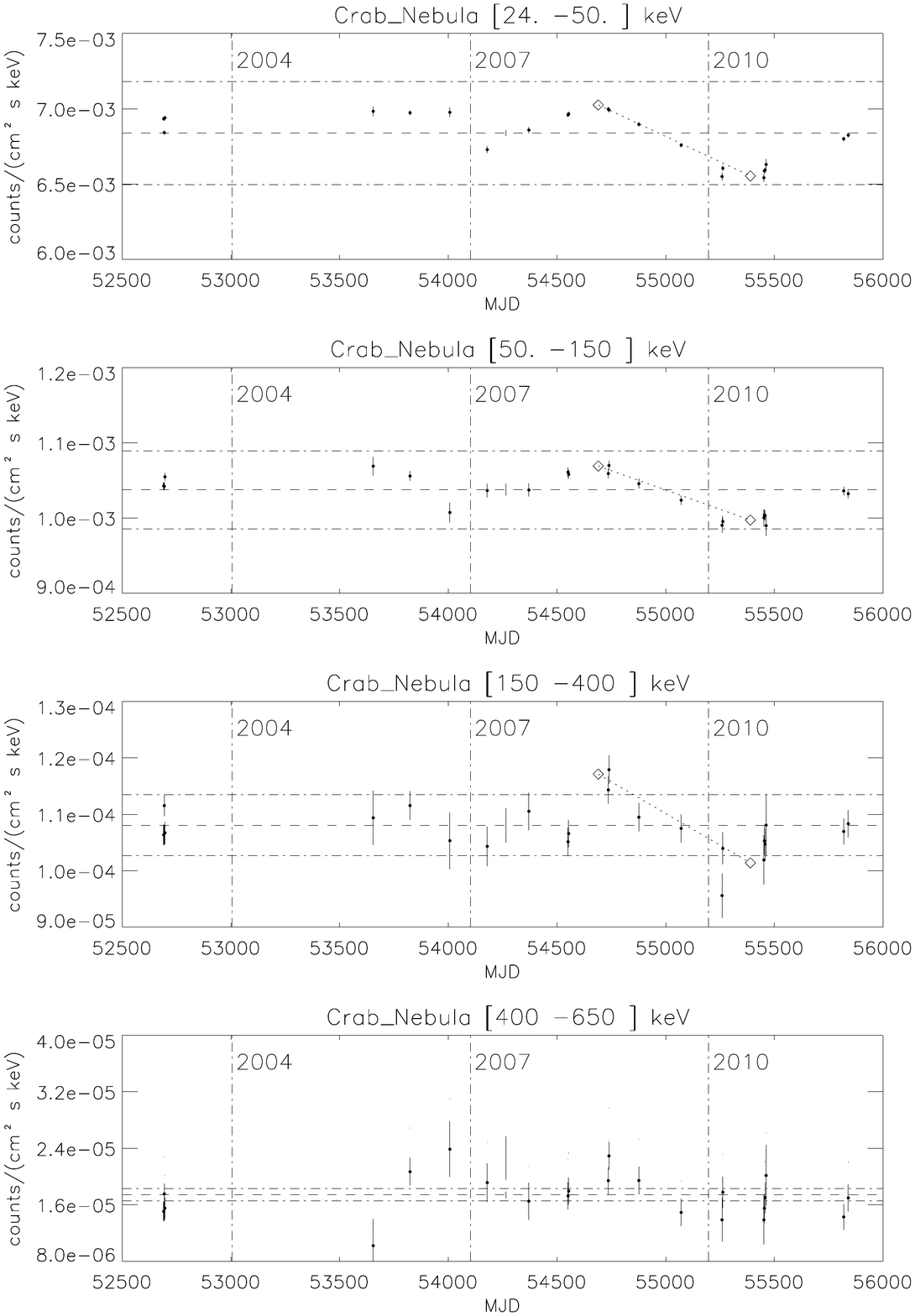}\includegraphics[angle=0,width=9cm,height=22cm,trim= 0.cm 0cm 0cm 2cm]{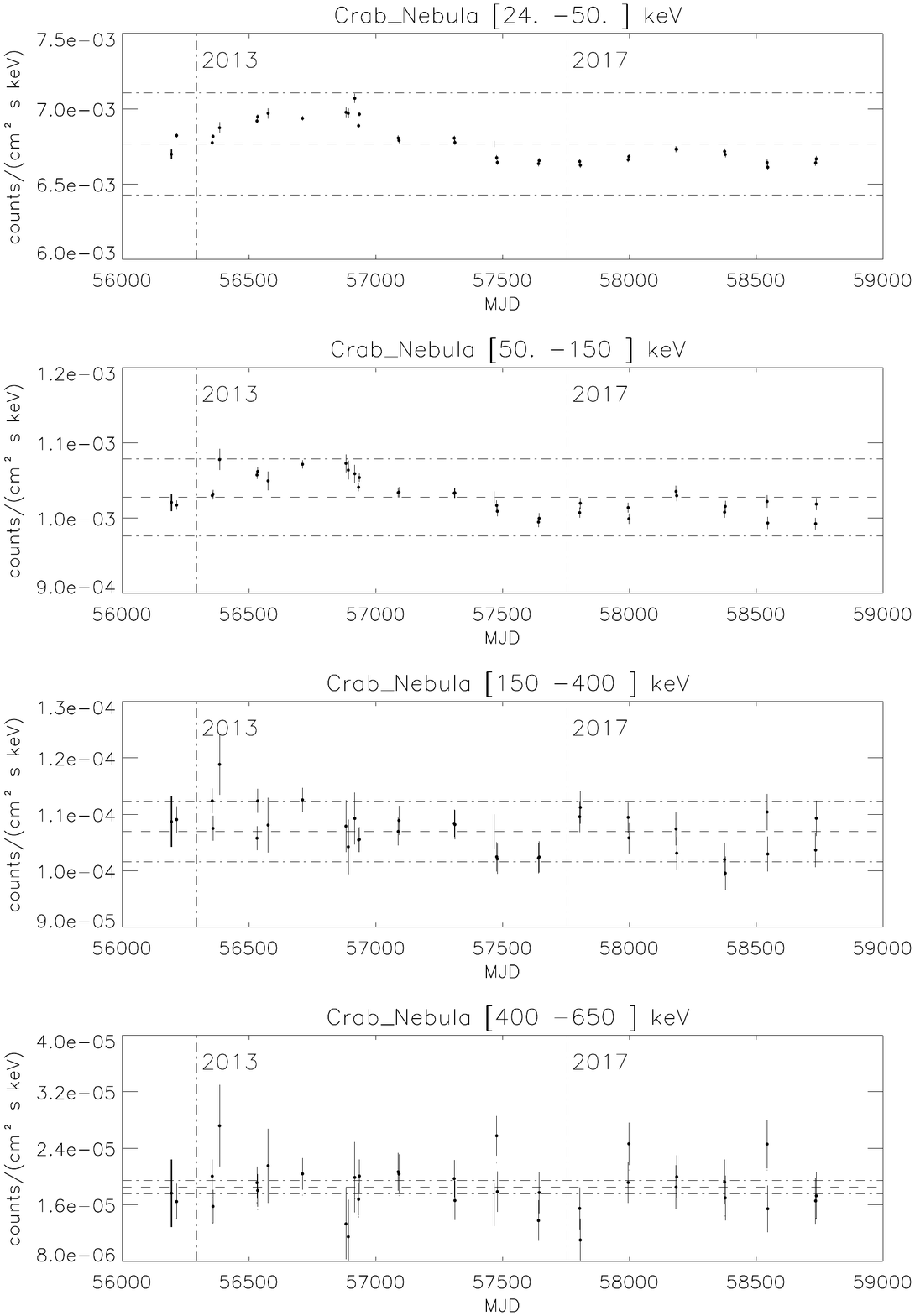}

\caption{Light curves of the Crab emission in four broad bands along the mission.
Each point corresponds to one revolution ($\sim$ 0.5 to 2.5 days of useful duration); the two panels correspond to the two periods mentionned in the text (before and after 2012 May 13); the dashed lines represent the mean flux, and the dot-dashed lines the $\pm$ 5\% brackets around it. Dotted lines stand  for the source decline reported by \cite{crabvar}, with a slope of 3.5\% yr$^{-1}$ from 24 to 150 keV and 7\% yr$^{-1}$ for the 150-400 keV band.}\label{CLrev}.
\end{figure}

A  synthetic view of the Crab evolution is obtained by looking at  the source flux averaged by revolutions (see Figure \ref{CLrev}) in the four broad bands.
Depending on the scheduled planning, the useful durations of the exposure dedicated to the Crab Nebula in the individual revolutions, range from 20 to $\sim$ 100 scw, i.e. 40 to 230 ks.\\

From these plots, it appears that the long term evolution of the Crab is slowly varying around a stable mean flux. The variability level in the range 24-150 keV is of the order of 5\%, represented by the two dotted lines in Figure \ref{CLrev}. This remains mostly true for the 150-400 keV energy band even though higher variations are  sometimes observed. The most important evolution is that reported by \cite{crabvar} in 2008-2010.  
A flux decline has been measured by several instruments, implying it belongs to the source. It is characterized by a linear decrease of $\sim$3.5\% yr$^{-1}$ between 15 and 100 keV and roughly twice this value between 100 and 300 keV, for the period MJD 54690-55390. These variations are illustrated by a dotted line in Fig.\ref{CLrev}, in good agreement with the SPI data. The rms of the residual deviations around these templates present rms values of 0.6, 1.1 and 3.7\% for the three first energy bands respectively, which can be used as upper limits on the SPI efficiency uncertainties. 
In 2015, the source recovers  the same flux level as in 2009. This may be taken as a strong indication that the instrument efficiency is stable up to that time.  
For the lastest available observations, the Crab emission
exhibits a behavior similar to the 2008-2010 episod, but remains below the long term mean, yet within the same range ($<$ 5\%) as previously measured. However, in this case, the observed trend can be attributed  to both source or instrument evolution.
In the highest energy band (400-650 keV), all fluxes are compatible within 2 $\sigma$, and 
 the limited signal to noise ratios prevent any firm conclusion.
 A study of the instrument efficiency at higher energy (1.5 MeV) is presented in section \ref{40K} by means of the long term evolution of a specific background line.\\
 
To look in more details at the source evolution, fluxes have been extracted for each INTEGRAL exposure,  which lasts generally  $\sim$ 0.5 to 1 hr.

\begin{figure}[ht]
\includegraphics[width=9cm,height=\textheight]{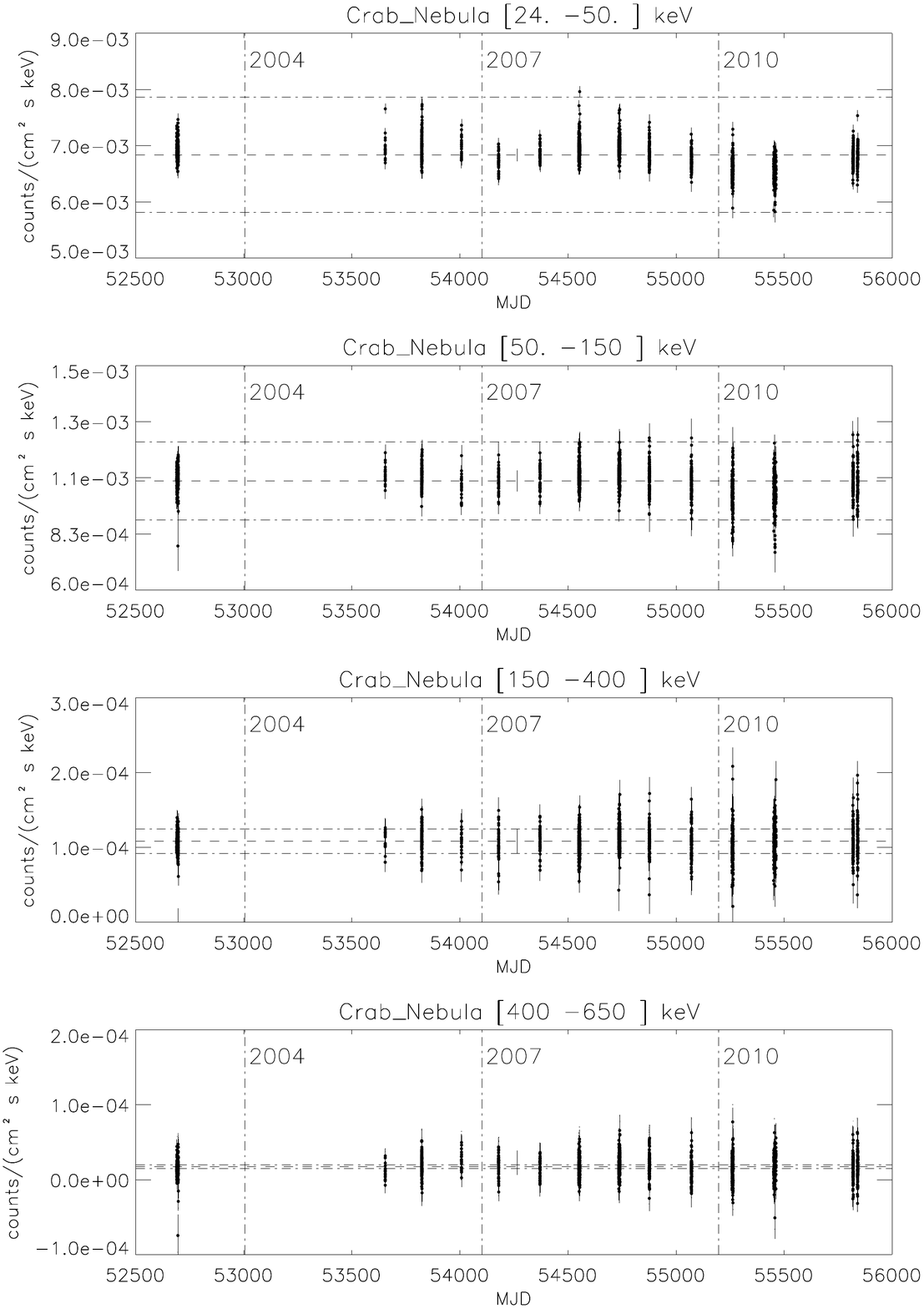}
\includegraphics[width=9cm,height=\textheight]{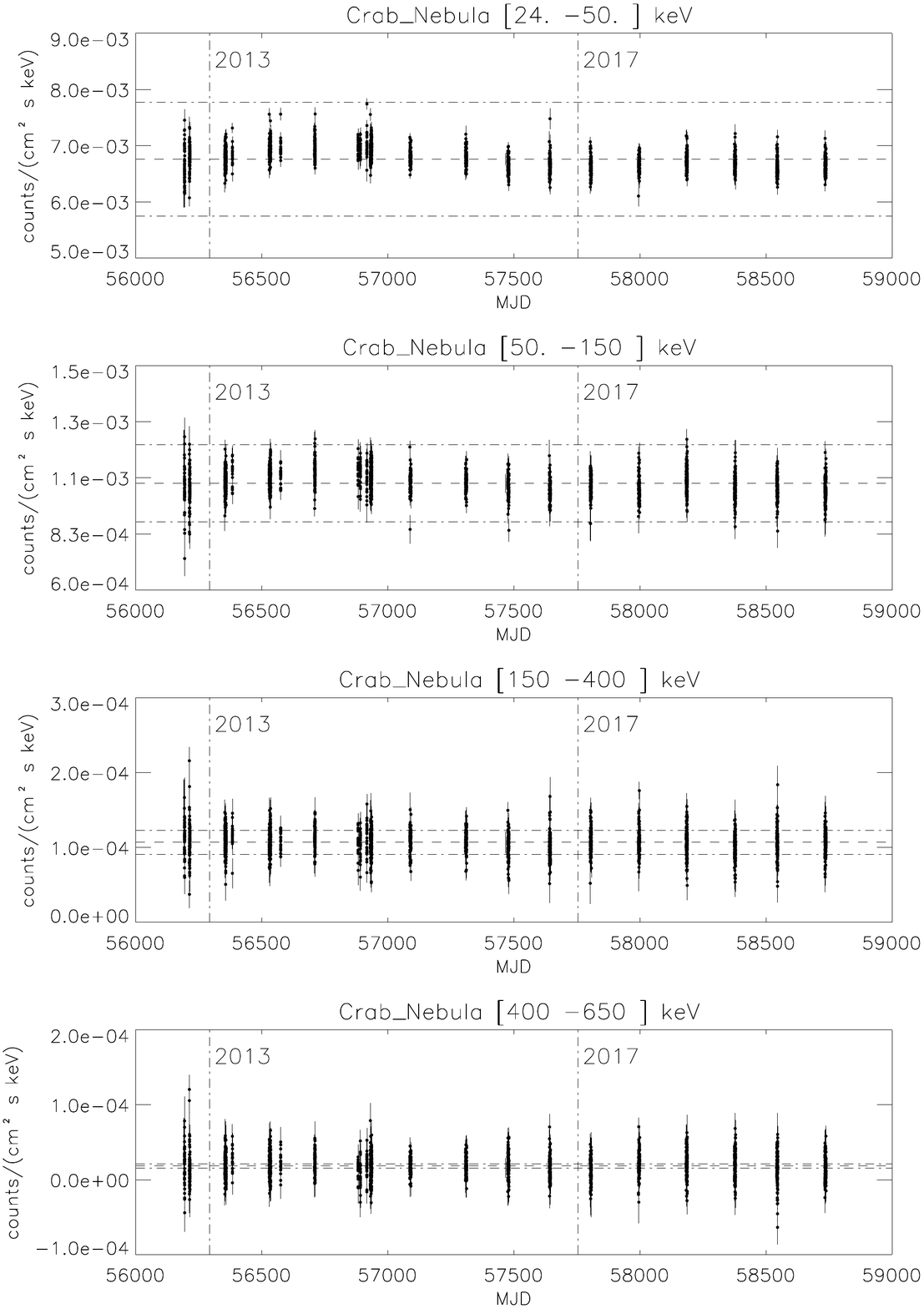}

\caption{The same as Fig. \ref{CLrev} but with each point corresponding to one scw ($\sim$ 0.5-1 hr) and the dot-dashed lines bracketing $\pm$ 15\% around the mean values.}\label{CLscw}.
\end{figure}

\begin{figure}[ht]
\includegraphics[width=5.5cm,height=5.cm,trim= 1.cm 13cm 1cm 3cm]{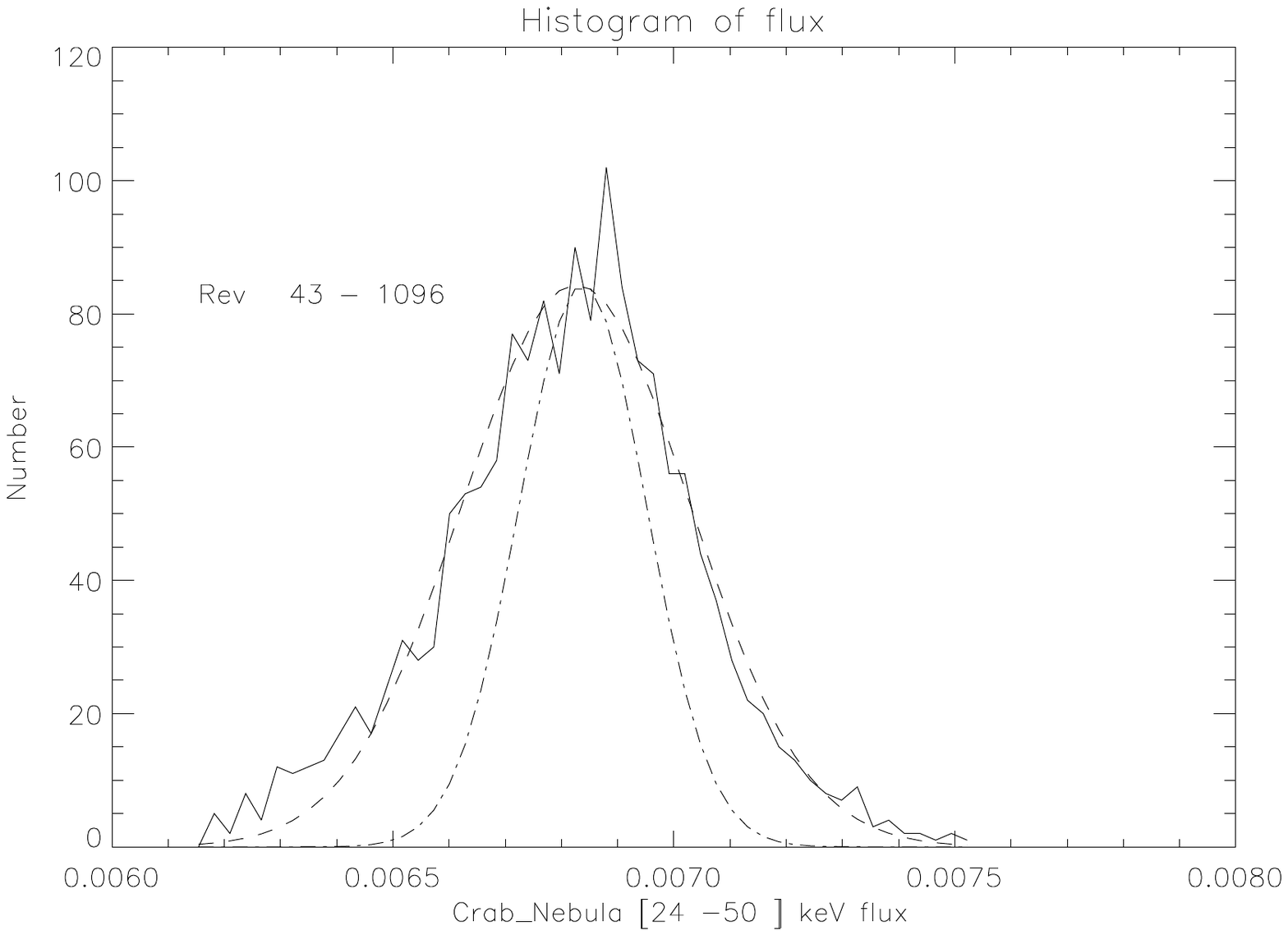}\includegraphics[width=5.5cm,height=5.cm,trim= 1.cm 13cm 1cm 3cm]{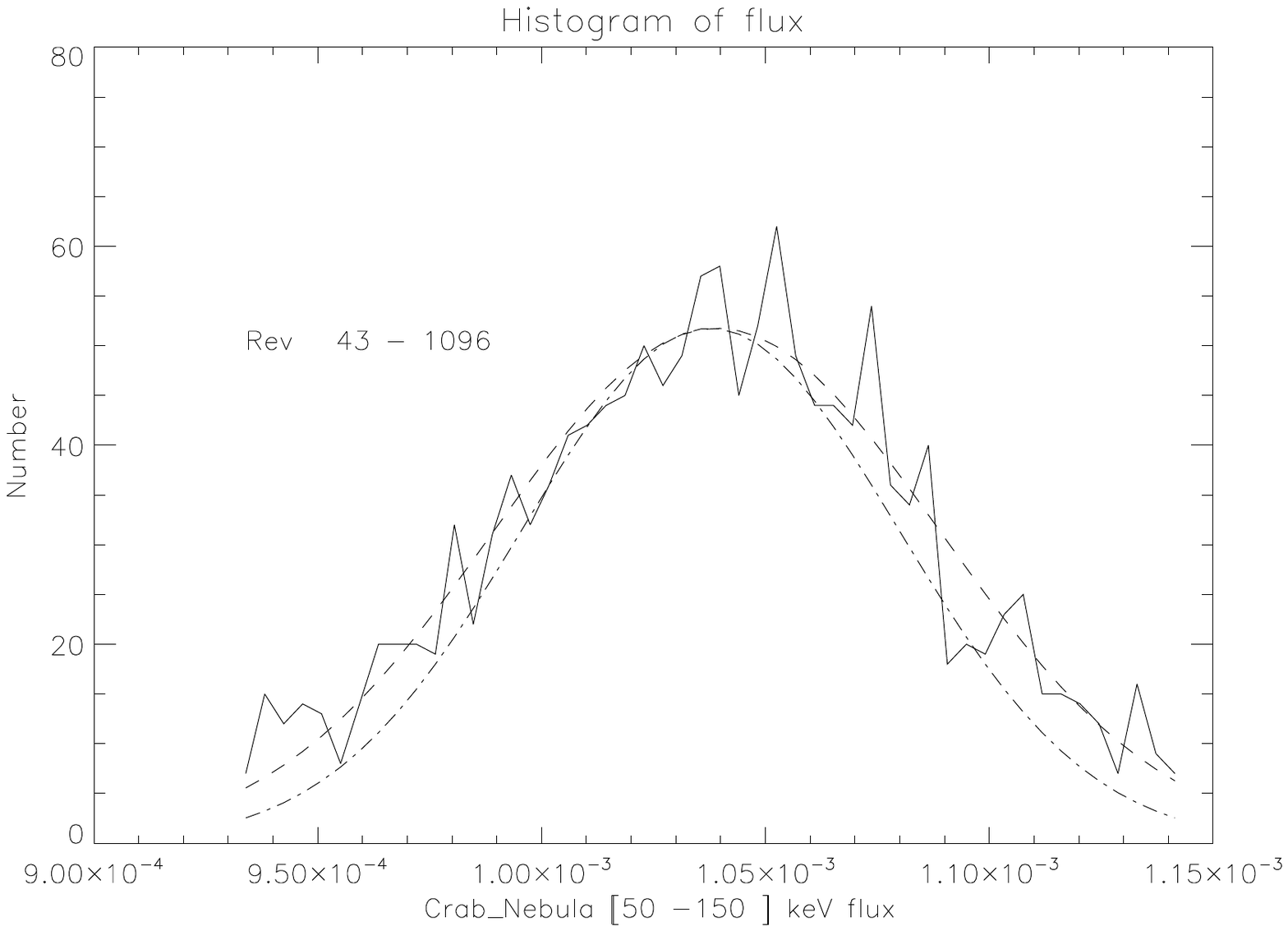}
\includegraphics[width=5.5cm,height=5.cm,trim= 5.cm 13cm -2.5cm 3cm]{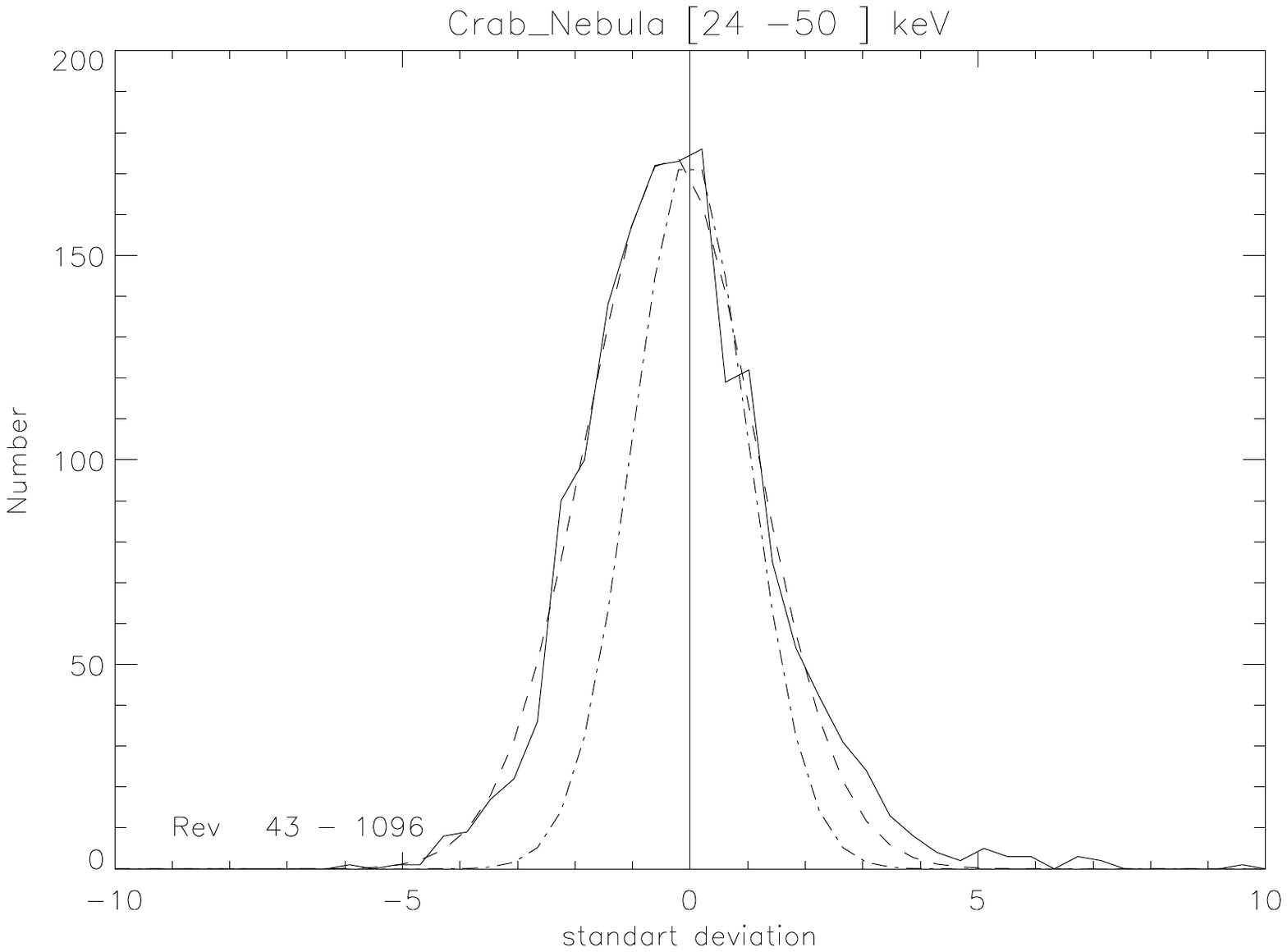}
\includegraphics[width=5.5cm,height=5.cm,trim= 1.cm 13cm 1cm 3cm]{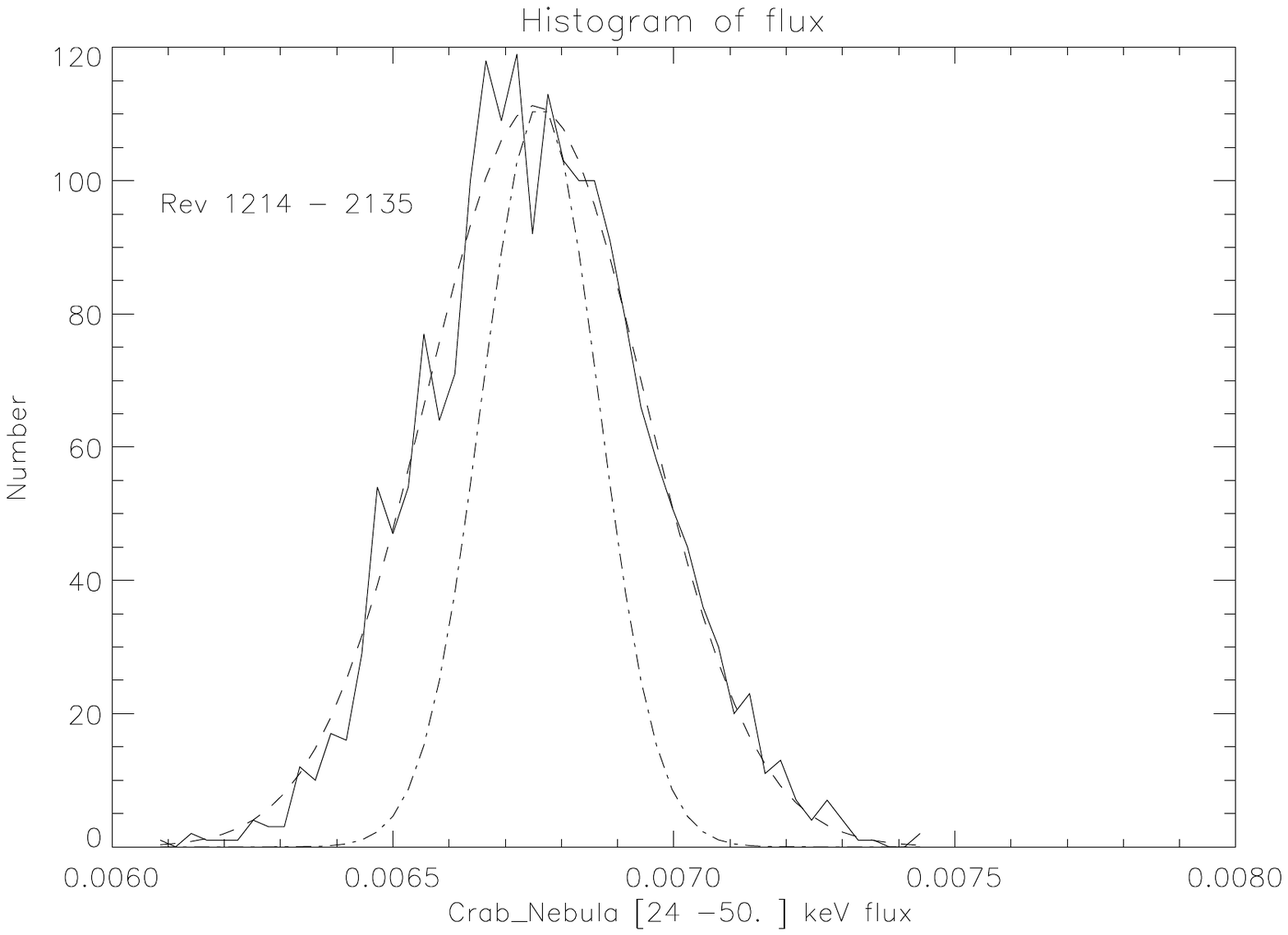}
\includegraphics[width=5.5cm,height=5.cm,trim= 1.cm 13cm 1cm 3cm]{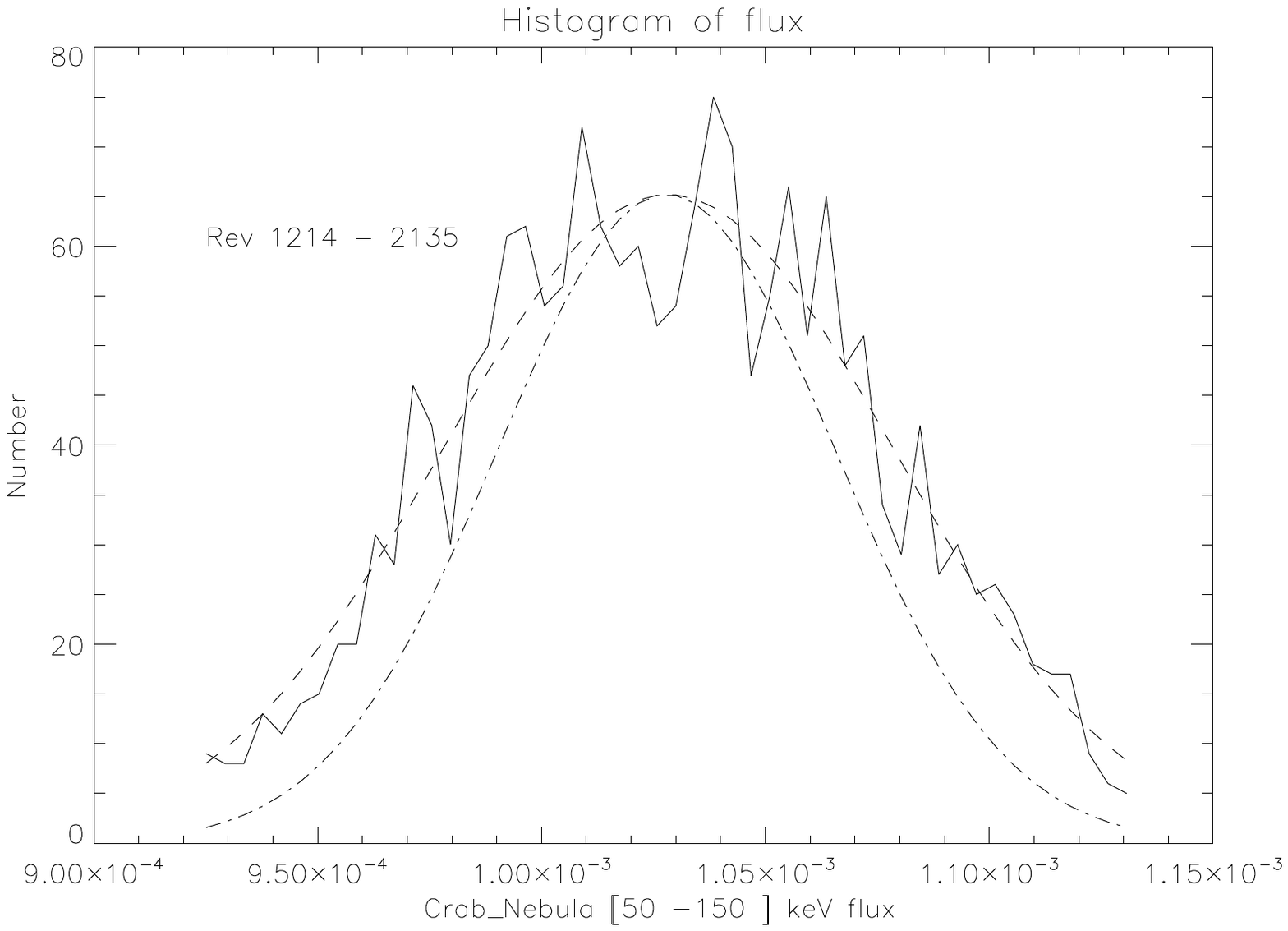}
\includegraphics[width=5.5cm,height=5.cm,trim= 0.cm 13cm 1cm 3cm]{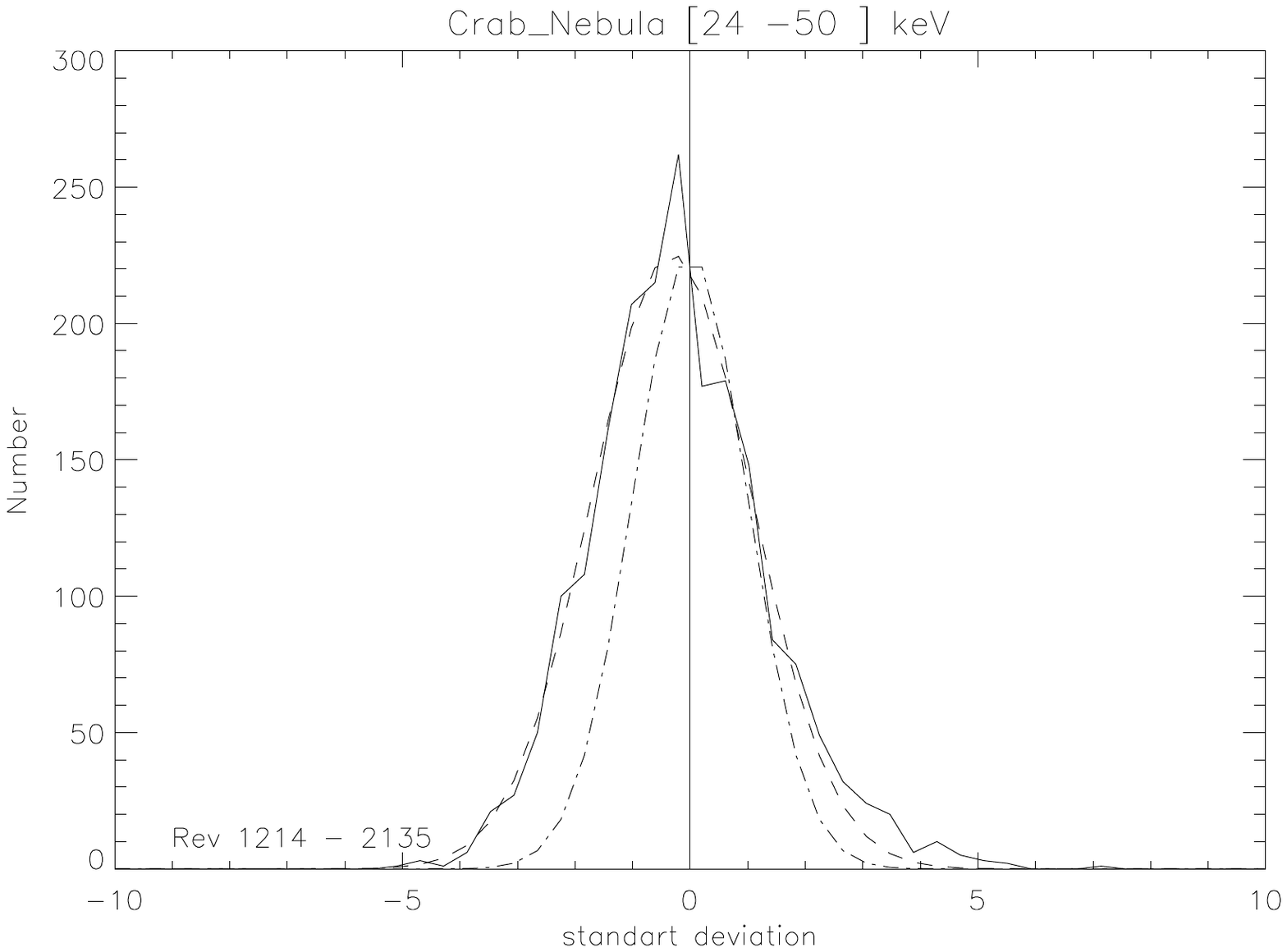}
\caption{Histograms of the measured fluxes (scw timescale) for each subdataset (top and bottom resp.).\\
 Left: 24-50 keV energy range; middle: 50-150 keV one; right: histograms of the flux measured in units of standart deviations, for the 24-50 keV energy range. The solid lines correspond to the data (see text for details), the dashed lines to the associated best fit normal laws and the dot-dashed lines to the distribution expected from the observed means  and errors.}\label{histo}.
\end{figure}

At this timescale, the source stability is less constrained, due to larger measurement uncertainties. Figure \ref{CLscw}  shows that between 24 and 150 keV (the most significant results, because the highest signal to noise ratios), the source flux dispersion is $\pm$ 15 \% (dot-dashed lines) around the  mean value (dashed line) or even less within each revolution.
Several effects contribute to the observed dispersion: statistical error on the measurements, instrumental systematics and intrinsic source variability.
 To investigate this point, the histograms of the flux measurements have been built for the first two bands (24-50 keV and 50-150 keV) and shown in  Figure \ref{histo} (solid line).

The expected dispersion around the observed mean flux ($F_{mean}$), due to measurement uncertainties, is  estimated from the mean value of the statistical errors  (see Table \ref{val}). The corresponding normal laws are plotted in dot-dashed line. At the same time, the observed distributions are broader than the purely statistic ones. They can be  described by normal laws, whose parameters (identified as $\mu$  and $\sigma$) are also given in Table \ref{val}. From the comparison of observed and expected values, we deduce that the broadening of the observed distributions, in all cases, corresponds to  an additional variance, attributable to instrumental systematics and/or intrinsic source variability,  equal to $\sim$ 2.5 to 3.3\% of the mean source flux. We note also that 86\% (for the first period) and 91\% (for the second one) of the measured fluxes on the scw timescale are within $\pm$ 5\% of the respective mean fluxes in the 24-50 keV band. These values become 67 and 69\%, in the 50-150 keV band.\\
 However, the source flux has been shown to vary by a few \% on a few year timescales. This must be taken into account to access more rigorously to its hour timescale variability. We have thus considered the data in terms of standart deviations relative to the  flux average observed in the individual revolutions. To do that, we consider the difference between each scw flux and the average value measured over the revolution, normalized by the error.   The corresponding histograms, for the first energy band, are presented in the right panels of Figure \ref{histo}. While the statistical part of the dispersion is represented by a normal law (0,1),  the data follow a slightly shifted ($\mu$ around -0.35) and broader ($\sigma \sim$ 1.4) distribution (see Table \ref{val}). In both cases, the broadening of the observed dispersion corresponds to  an additional variance of  $\sim$ 1. Using the values of $F_{mean}$ and  $<Err>$ reported in Table \ref{val}, this results in a level of  variability,  due to instrumental systematics and/or intrinsic source behavior, ranging  between  1.5 and 1.8\% of the mean  source flux. This  quantifies the remarkable steadiness of the  Crab emission over periods of several days or weeks.\\
In the two higher energy bands, the dispersion reaches a much larger factor (up to a few) but is comparable to the measurement uncertainties on this short timescale. \\

The conclusion of this broad band analysis is twofold: first, the SPI  instrument efficiency loss (at least between 20 and 400 keV) is below 5\% over more than 17 years of operation;  and second, even if the origin of the measured variability remains to be assessed (intrinsic to the source, instrument systematics or both), the value is modest
enough to consider that the Crab emission is constant to  first order over decades, with a variability level
within $\pm$ 5\%, comparable to the instrumental uncertainties. Furthermore, when looking at timescales  of a few days or weeks, the upper limit on source variability or instrument uncertainties is below 2\%.

\subsection{spectral analysis}
\begin{figure}[ht]
 \includegraphics[width=8.5cm,height=10cm,trim= 0cm  5cm 0cm 0cm]{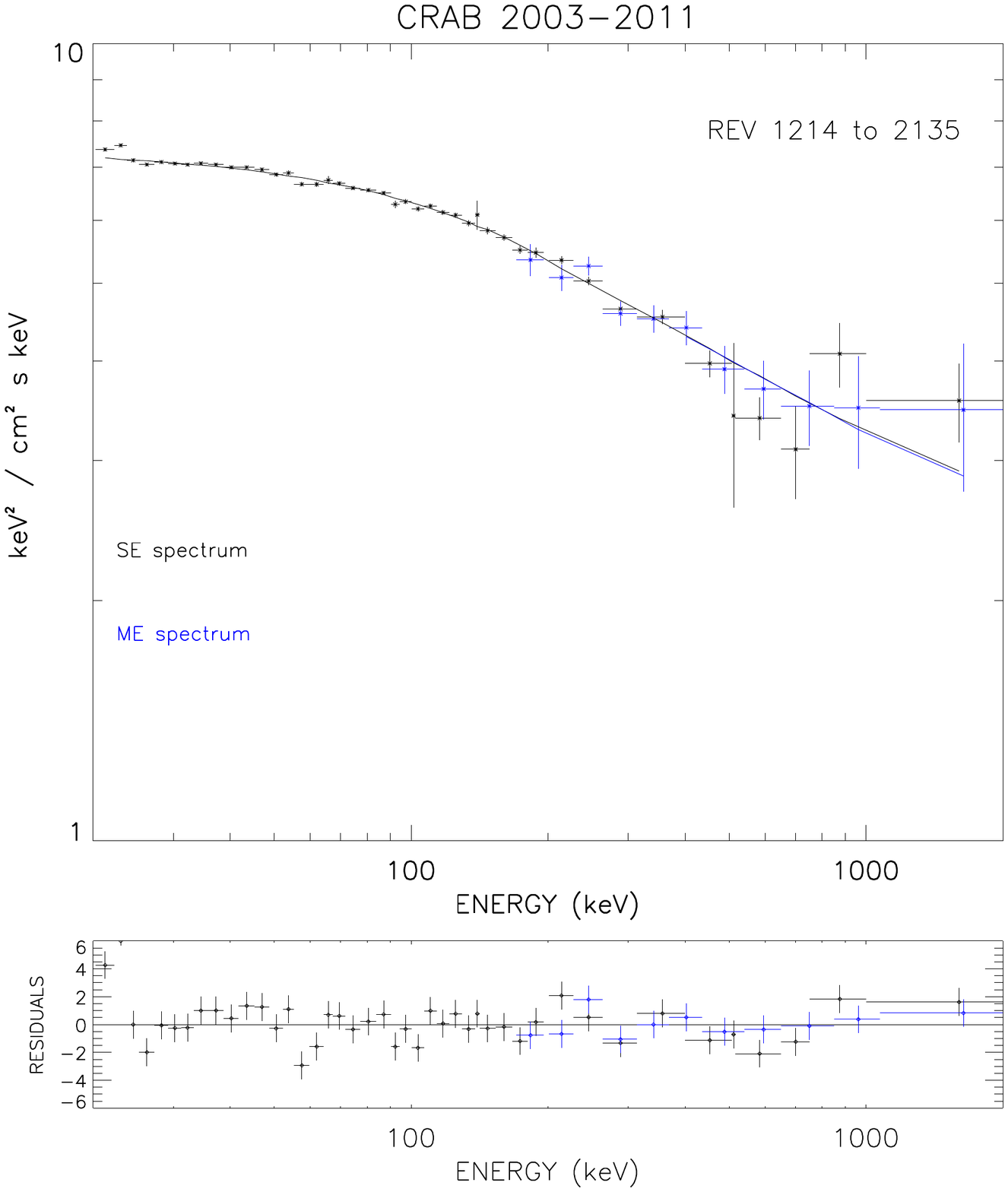}
 \includegraphics[width=8.5cm,height=10cm,trim= 1.5cm  5cm -1.5cm 0cm]{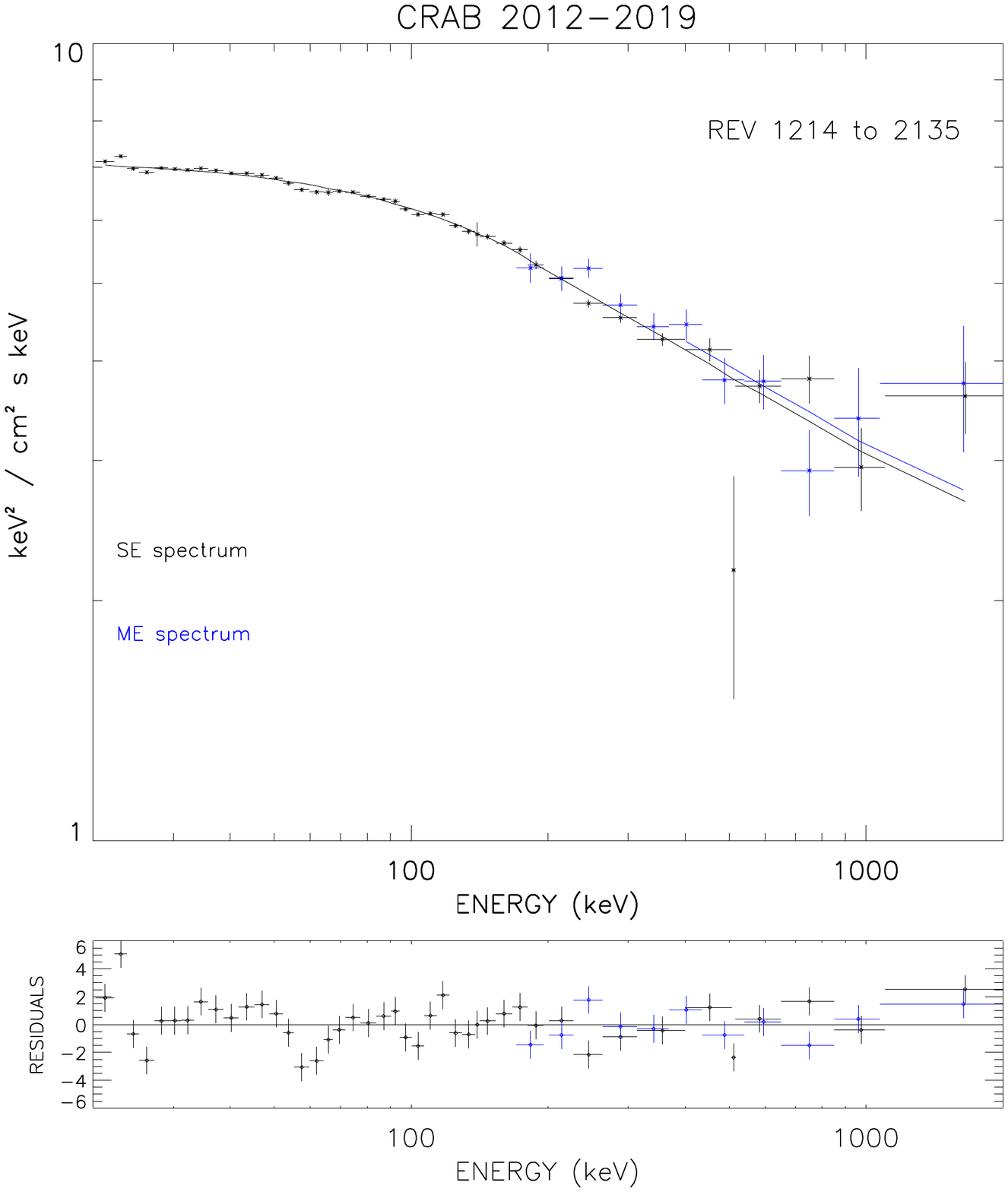}
\caption{Spectra extracted from one-detector events (black points) and double-detector events (blue points) for the two considered periods (see text and Table \ref{fit}).} \label{SP}
 \end{figure}

The Crab Nebula energy spectrum is often described by a power law for most of the energy domains covered by individual instruments. However, the observed slopes are  energy (or instrument) dependent,
 indicating a more complex spectral shape when considering the whole electromagnetic emission (see for instance \cite{crabComptel} and references therein). More specifically, the hard X-ray domain covered by the SPI instrument makes the link  between the X-ray domain (observed slope around 2-2.1 \citet{kirsch05}) and the $\gamma$-ray domain (slope of 2.23, \citet{crabComptel}).   
 A broken power law was long used to described the slope evolution, with a break around 100 keV (e. g.  \cite{crabNRL,crabBatse}). A single curved shape with a slope varying with  Log(E) has then been proposed \citep{crabMassaro, mineo06}, but this model cannot be extrapolated in a broader energy domain, due to the continous curvature. To avoid an unphysical break while allowing extrapolations toward lower and higher energies, we proposed to  describe the broad band spectral emission by the Band model, used by \citet{band93} for modeling GRBs emission (GRBM in Xspec language), which joins two power laws by a smooth curvature.
  This shape better reproduces the smooth slope evolution observed between 20 keV and $\sim$ 1 MeV than previously proposed models, even though the Band model remains an analytical description, without physical law behind.\\
 We start our study by   analyzing the  mean spectra, obtained by summing
 the spectra belonging to the two periods defined above. We fit together
 the one-detector and two-detector event spectra for both periods, imposing the
 same  shape parameters while the 2 normalizations are let free (Figure \ref{SP}). 0.5\% of systematic has been added during the fit procedure with Xspec.
 The best-fit parameters  are reported in Table \ref{fit} together with those obtained for the sum of all the observations. 
 
 Table \ref{fit} contains also the 30-100 keV flux, which is more representative of the source flux than the best-fit normalization parameter, because independent of the shape parameter values.    
The parameters of both datasets are in good agreement. Rather than claiming true parameters values (which depend on the SPI calibration systematics), this result assesses the stability of the source spectral shape on long term scales, and again, that of the instrument response.\\

\begin{figure}[ht] 
  \includegraphics[width=8.5cm,height=18cm,clip,trim=1.cm 0.cm 9.5cm 0.cm]{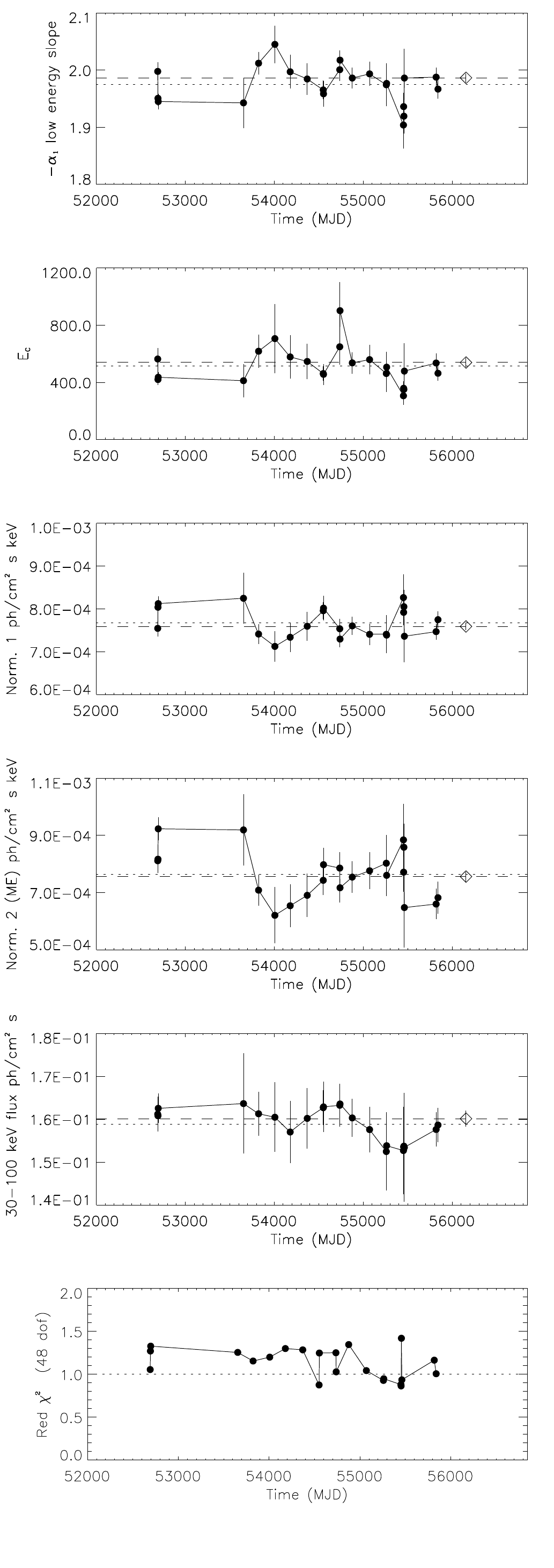}
  \includegraphics[width=8.5cm,height=18cm,clip,trim=1.cm 0.cm 9.5cm 0.cm]{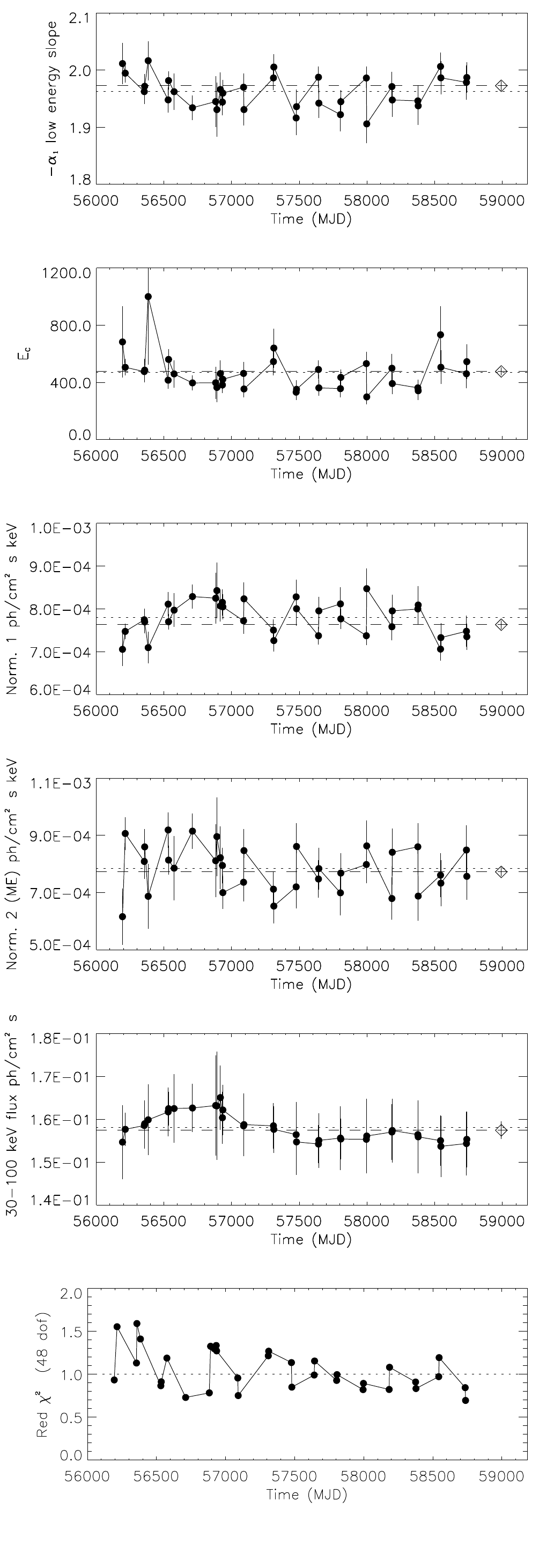}  
\caption{ Best fit parameter evolution with time. In the GRBM model, $\alpha$ represents the low energy slope and $E_c$, the cutoff characteristic energy. The high energy slope has been fixed to 2.3. N1 and N2 correspond to the model normalization for the one-detector event  and two-detector event  spectra, resp.. The dot-lines trace the mean values (averaged over the panel dataset, except for the $\chi^2$ panels) and diamonds (+ dashed lines) stand for the averaged spectra values. 0.5\% of systematic has been added during the fit procedure.} \label{param}
 \end{figure}

To study the evolution of the spectral parameters with time, we have repeated the fit procedure for each individual revolution. To limit the degeneracy  between the parameters, in particular between the second slope and the characteristic energy, the 2nd power law has been fixed at the value obtained on the long term averaged spectrum (2.3) and compatible with the slope observed  above a few MeV \citep{crabComptel}. This makes the comparison of the parameters along the time more informative. \\
The best fit values of the remaining four free parameters are presented in Figure \ref{param}, together with the $\chi^2$ values and the 30-100 keV flux, versus time.\\

 The narrow ranges span by the low energy slope and the source 30-100 keV flux quantify the stability of the Crab emission, in terms of both shape and intensity. 
The characteristic energy (which drives the slope change) is less constrained. The two mean values are  only marginally compatible (see Table \ref{fit}). Even though this may point out a true evolution of the source emission above  a few hundreds of keV, we cannot exclude a decrease of the instrument efficiency at high energy (see section \ref{40K}).  

The normalization factors are driven by the low energy slope best fit paramater value, and are not useful by themselves. However, we note that the normalization factors of the ME2 (for E $\simeq$ 170 keV) and one-detector event spectra agree  within 10\% or less  in any period. This demonstrates the coherence and validity of the data analysis procedure for both kinds of events, including the flux corrections described in \ref{DataAn}.
    
\section{The 511 keV region}\label{511}
\begin{figure}[h]
 \includegraphics[width=8.5cm,height=3cm,trim= 1cm 13cm 1cm 6cm]{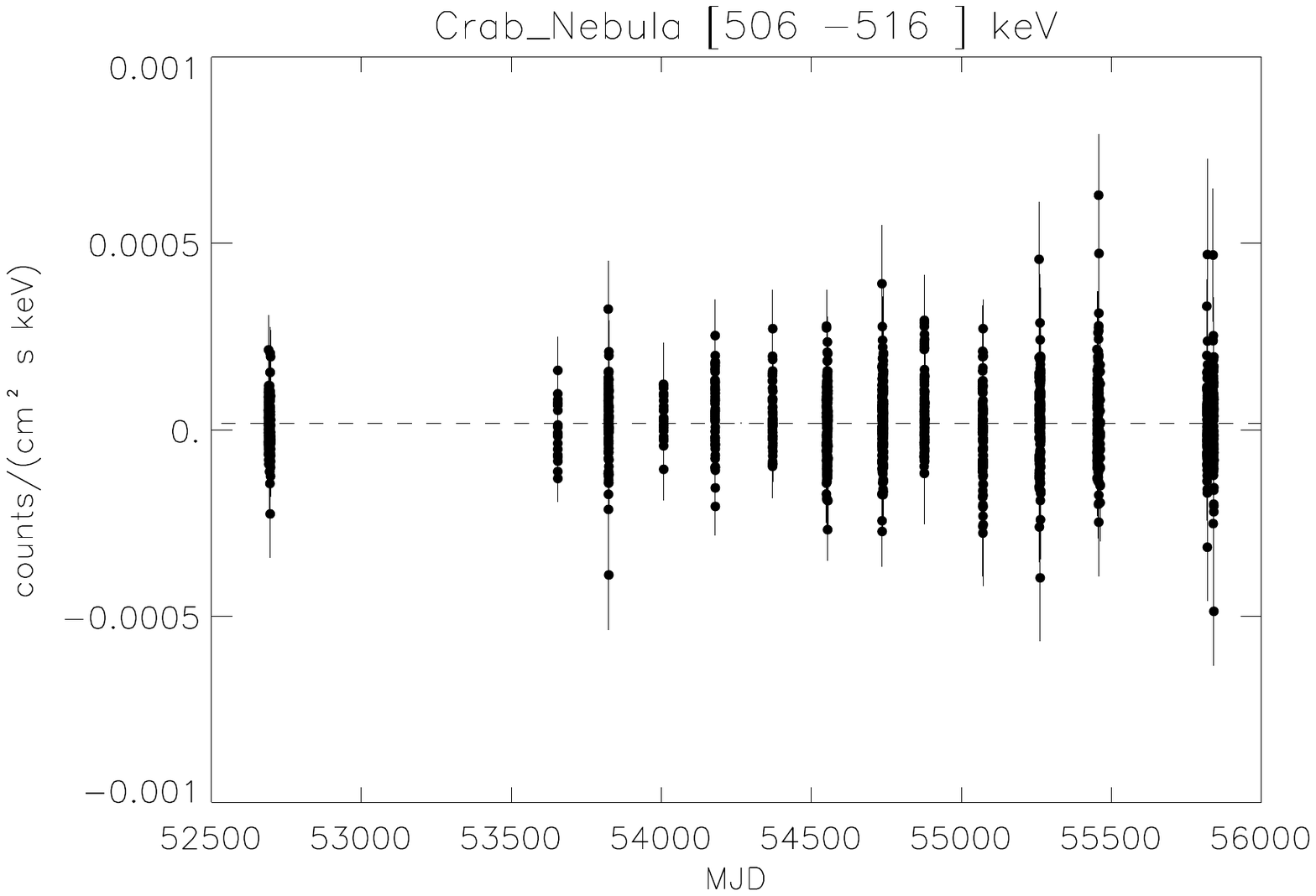} 
 \includegraphics[width=8.5cm,height=3cm,trim= 3cm 13cm -1cm 6cm]{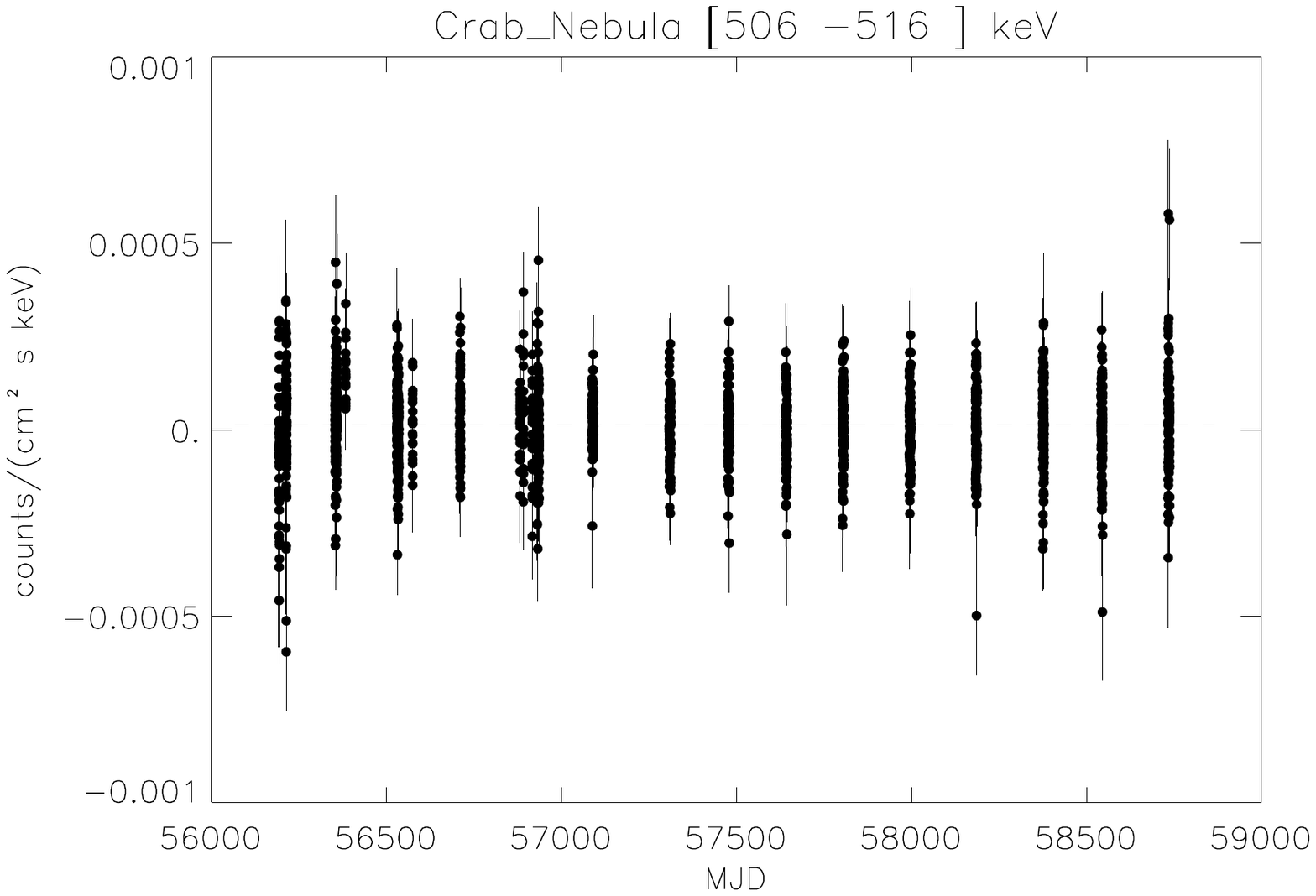} 
 \includegraphics[width=8.5cm,height=3cm,trim= 1cm 13cm 1cm 3cm]{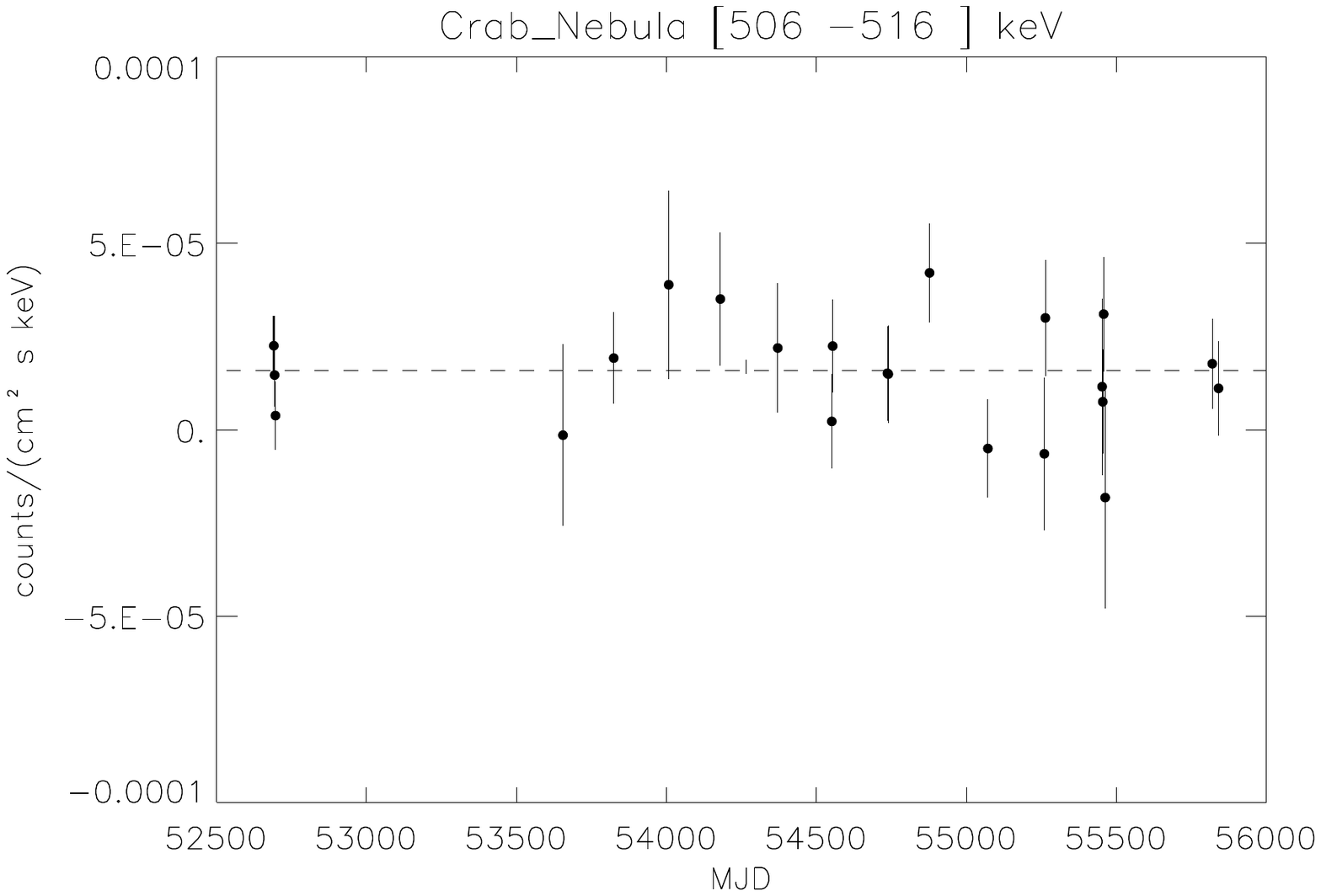} 
  \includegraphics[width=8.5cm,height=3cm,trim= 1cm 13cm 1cm 3cm]{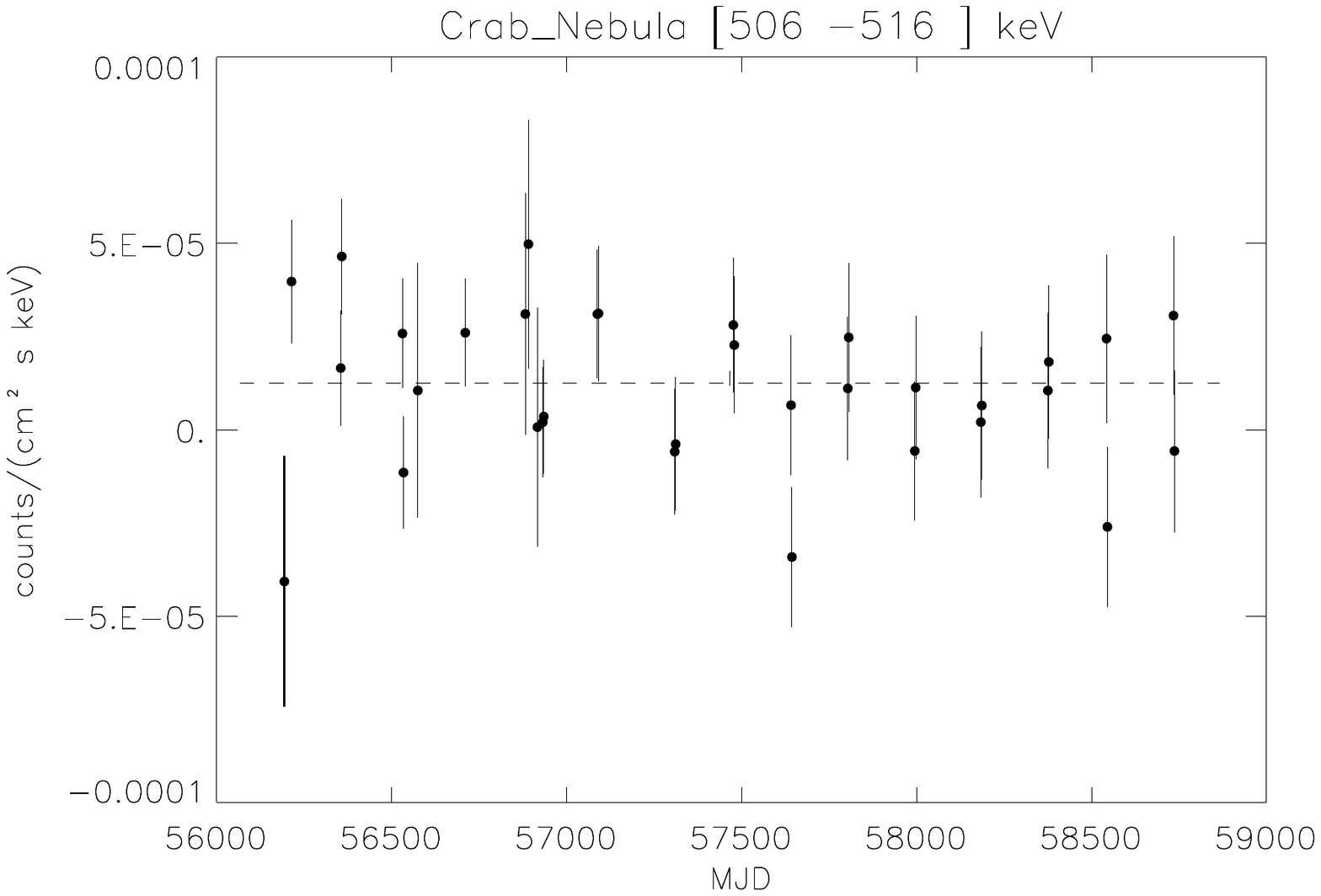} 
\caption{Top: Evolution of the Crab flux in a narrow  band (506-516 keV) around the 511 keV line. Each point corresponds to 1 scw ($\sim$ 0.5-1 hr) timescale. Bottom: the same as above, with each point corresponding to 1 revolution ($\sim$ 0.5-2.5 days of useful duration). The dashed lines trace the mean values (averaged over the panel dataset).} \label{511CLNA}
 \end{figure}

\begin{figure}[h]
  \includegraphics[width=8.5cm,height=3cm,trim= 1.cm 13cm 1cm 6cm]{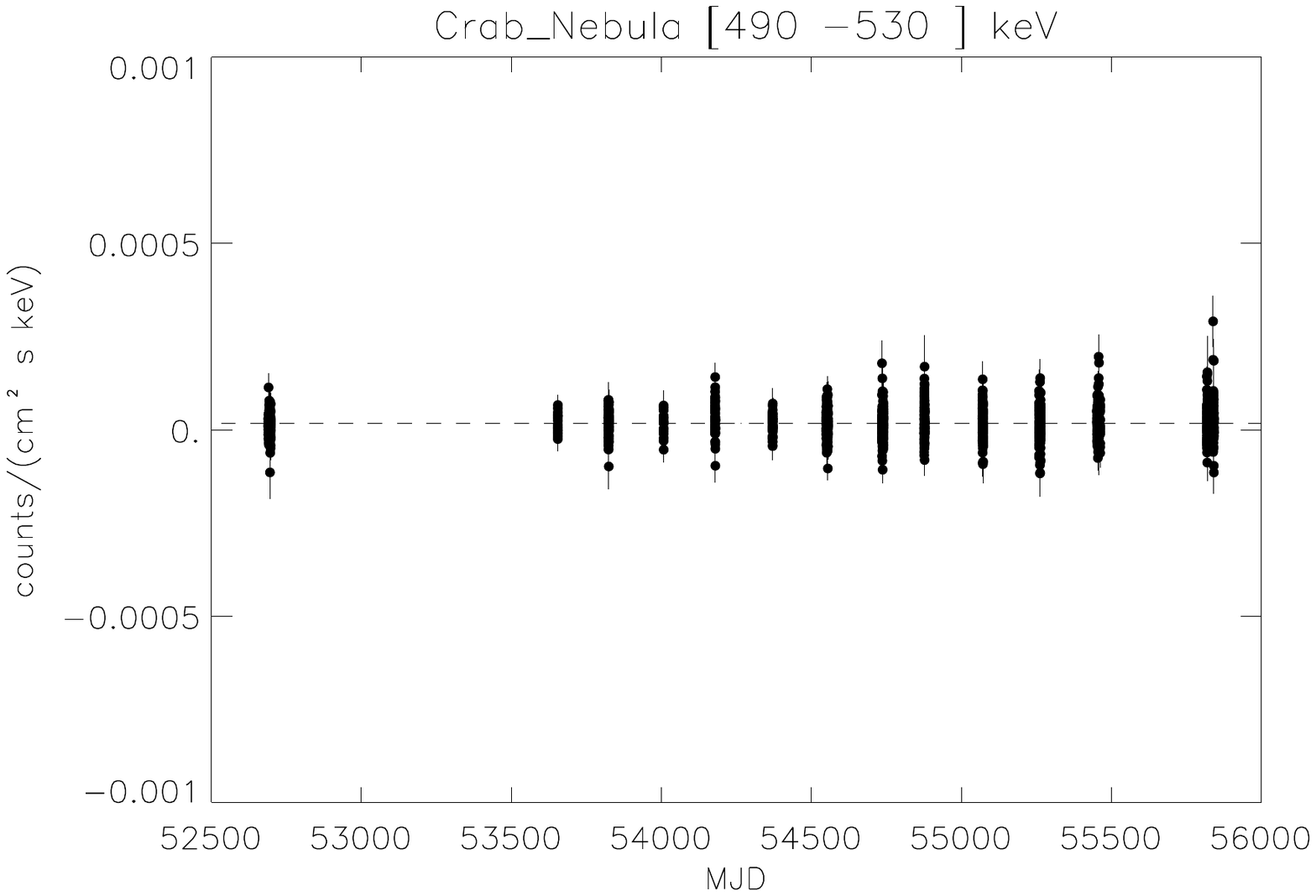} 
 \includegraphics[width=8.5cm,height=3cm,trim= 3.cm 13cm -1cm 6cm]{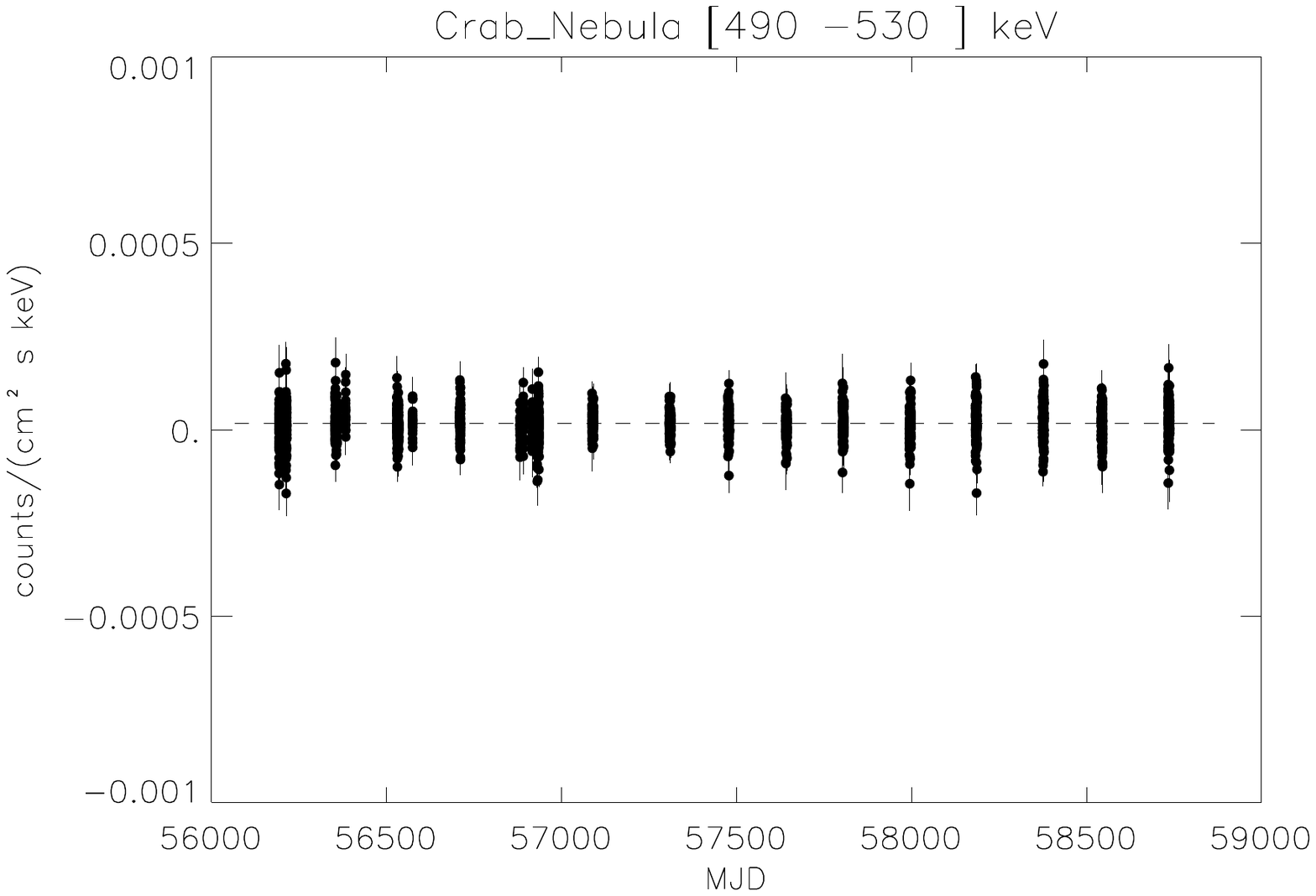} 
 \includegraphics[width=8.5cm,height=3cm,trim= 1.cm 13cm 1cm 3cm]{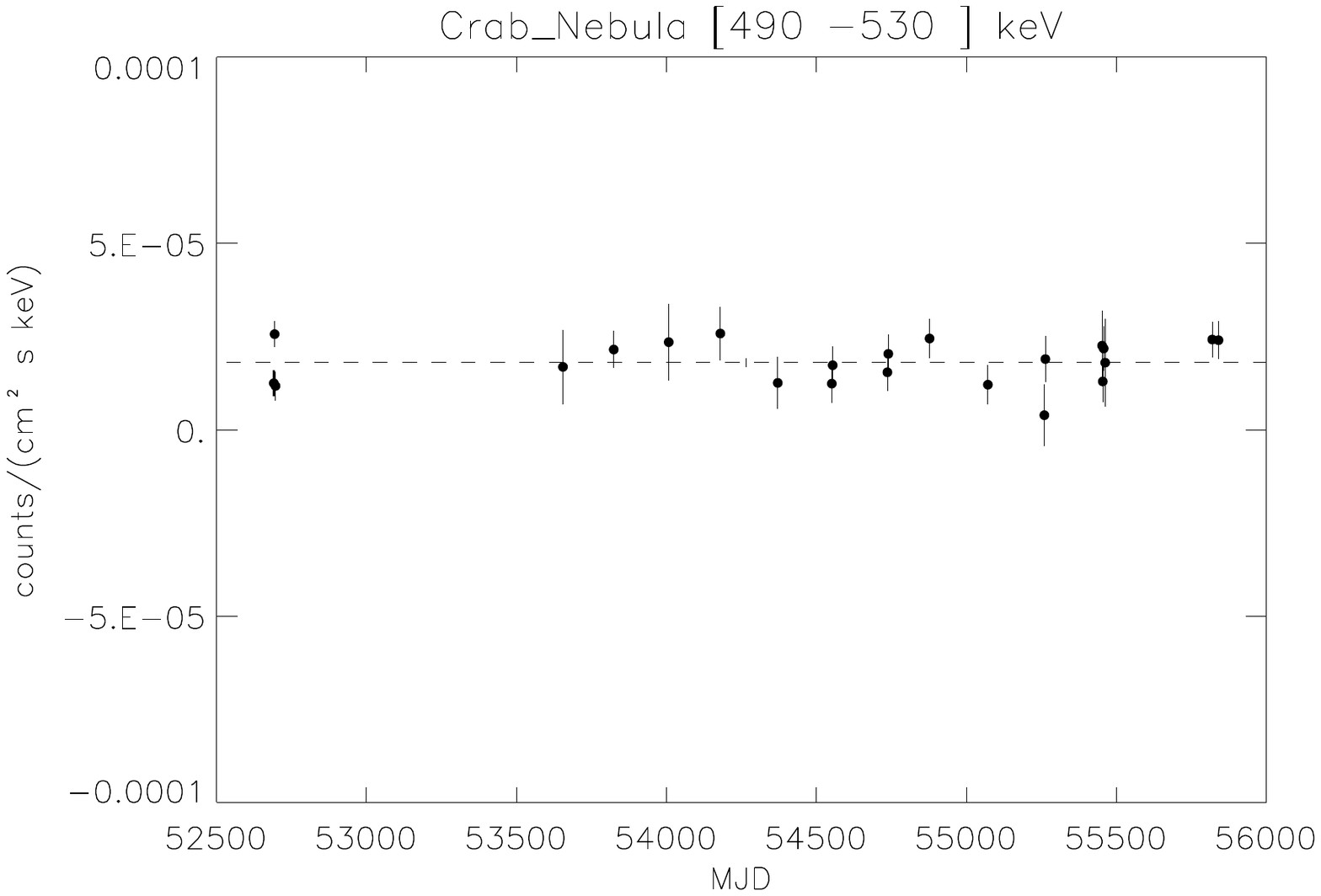} 
  \includegraphics[width=8.5cm,height=3cm,trim= 1.cm 13cm 1cm 3cm]{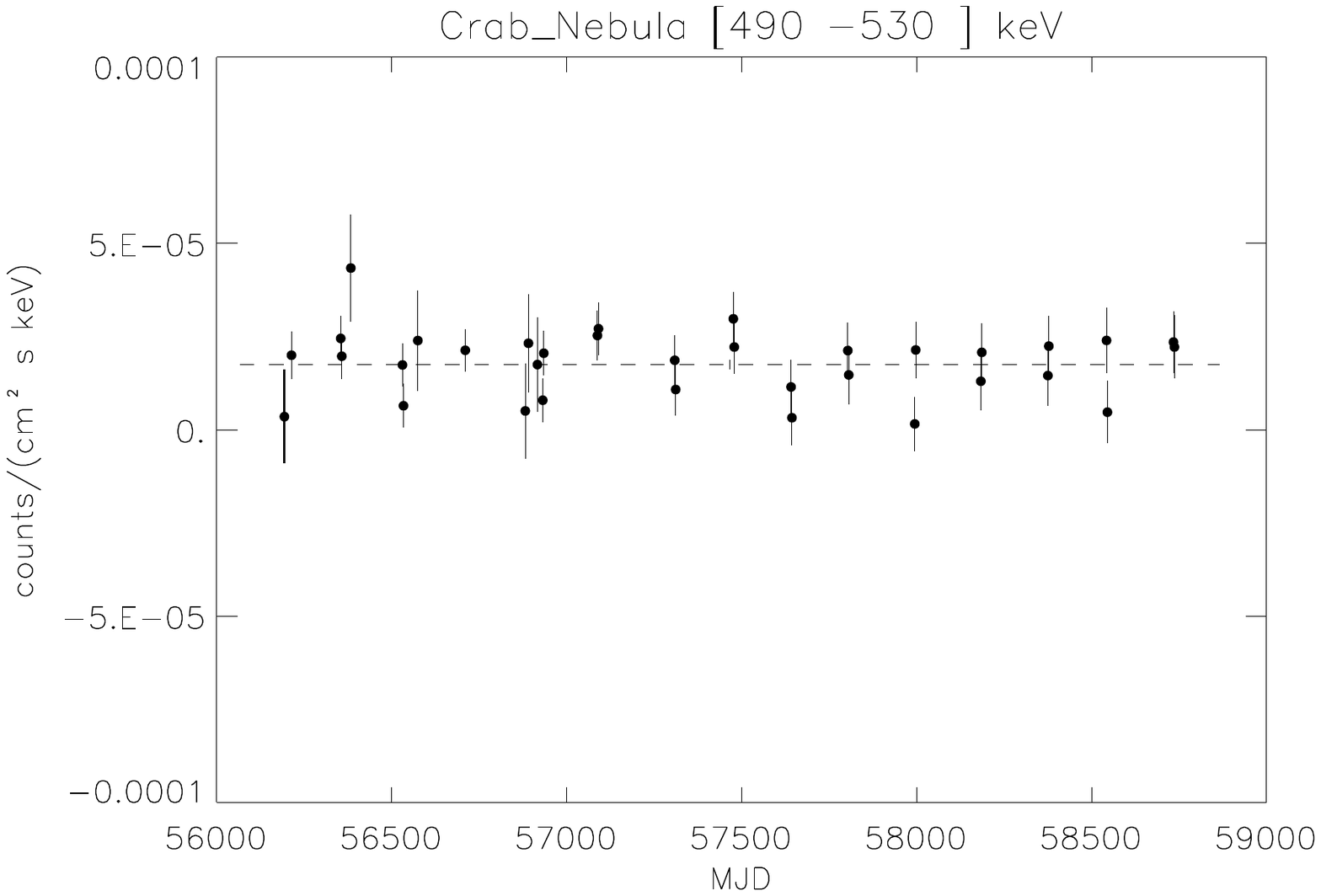} 
\caption{The same as Fig.\ref{511CLNA} for a broader energy band (490-530 keV).}\label{511CLBA}
 \end{figure}

The energy band around 500 keV is of particular interest for all $\gamma-ray$ emitters. It allows to test the presence of thermalized positrons in the emitting regions, and to get unique information on the particule distributions or other source parameters. The potential annihilation line may be modified by several physical factors (acceleration process, medium temperature and density...), impacting the observed central energy and width. Consequently, the access to the information is tricky. In order to scan  most of plausible
scenarii, we have extracted the source flux in two energy bands centered around the theoritical rest energy: a narrow (10 keV)  and a broad (40 keV) one. The resulting light curves (Figure \ref{511CLNA} and \ref{511CLBA}) exclude any transient emission on the hour and day timescales, disfavoring any annihilation sites inside the Nebula. The production of positrons cannot be excluded, but such particules had every chance to escape far away from the nebula before to slow down and annihilate.\\
 
 No detection was obtained for the line flux with  2 $\sigma$ upper limits of $2-4 \times 10^{-3}$ and $3-6 \times 10^{-4}$~ ph/cm$^2$~s in a 10 keV wide  energy bin 
 for the  hour and $\sim$ 1-2 days timescales.
  When a broader energy region is considered to take into account a potential line shift  and/or  broadening, the values become  $3-6 \times 10^{-3}$ and $5-7 \times 10^{-4}$~ph/cm$^2$~s respectively.
  For the total of the observations (7 Ms), upper limits on any annihilation feature are  $4 \times 10^{-5}$  and $6.5 \times 10^{-5}$~ph/cm$^2$~s, for a narrow and broad feature resp..
 
\section{SPI efficiency in the MeV range}\label{40K}

The steadiness of the Crab fluxes over the long term can be considered as evidence of the instrument stability within 5\% up to 400 keV. The low signal to noise ratio makes it difficult to extrapolate the conclusion  well beyond this limit.

However, the SPI efficiency at high energies can be affected by a drift of the Lithium ions from the central anode. This drift occurs during each annealing process. This phenomenon increases the effective radius of the anode and thus decreases the collecting volume in the Ge crystal. Consequently, the detector efficiency decreases in an energy dependent way. We have thus sought for a possible calibration source inside the instrument to test the highest energy range.
In the large sample of radioactive lines present in the SPI background spectrum (see the detailed study by \cite{weiden03}), we have identified one which  presents the suitable characteristics: the $^{40}K$ line at 1460.82 keV is produced by  natural radioactivity (ie 'the parent isotope is not radioactive because of activation processes in orbit', \citep{weiden03}), and its flux is high enough to be easily measurable on the revolution timescale. 
\begin{figure}[h]
 \includegraphics[width=18cm,height=6cm,trim= 1cm 13cm 1cm 3cm]{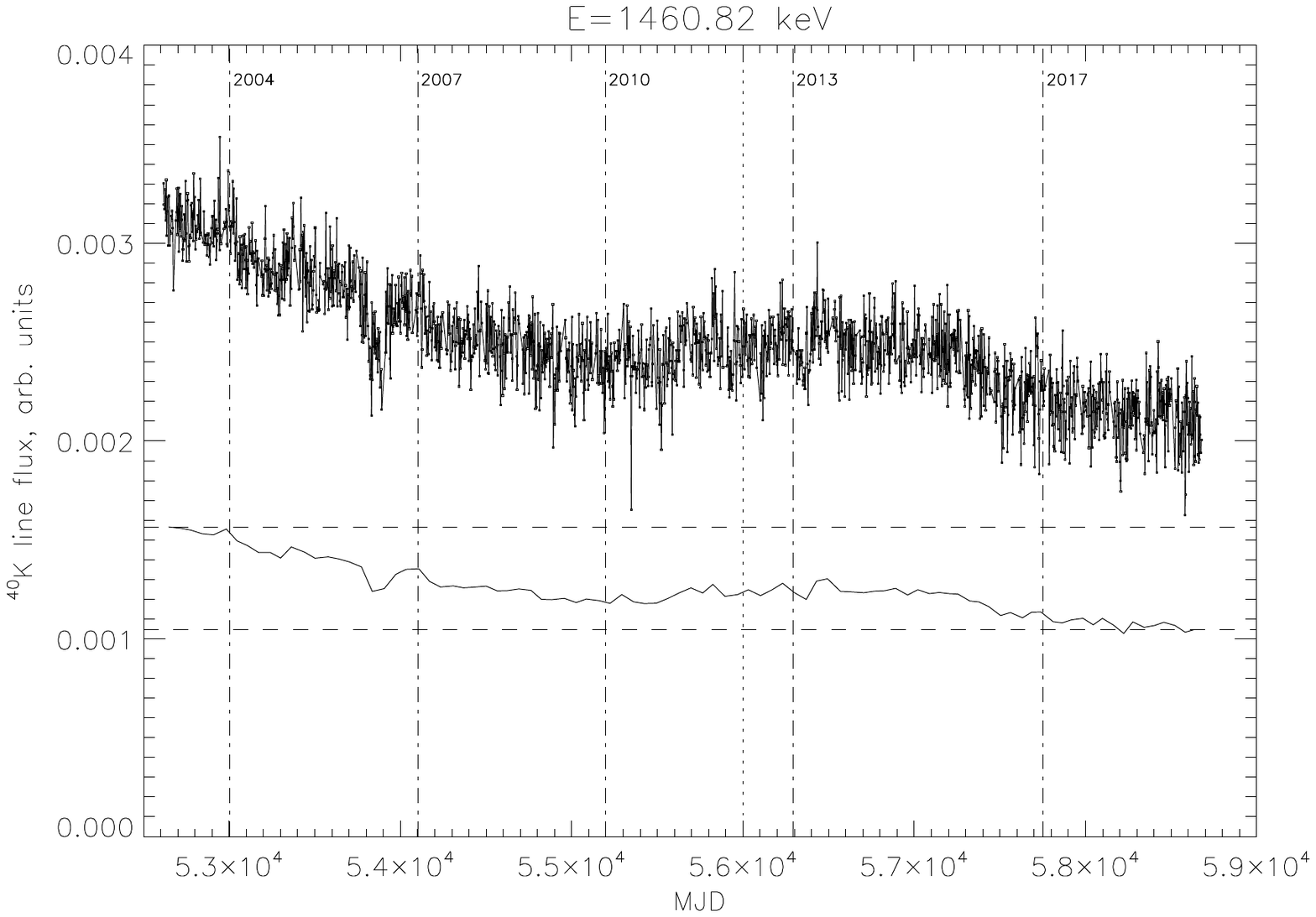} 
\caption{Evolution of flux contained in the $^{40}K$ line. The same data averaged over $\sim$ 2 months are displayed
below the original data (flux divided by 2 for clarity). The horizontal lines represent the first and last 2-month averaged values. The dotted vertical line at MJD=56000 demarcates the two subdatasets presented all along the paper.} \label{fig40K}
 \end{figure}
The determination of the $^{40}K$ line intensity requires a few precautions. In practice, we have considered, for all revolutions along the mission lifetime, the background spectrum, built with the PSD flagged events and recorded on the whole detector plane. In order to avoid a weak variable background line at 1463.95 keV (due to $^{72}Ga$), only the left half of the $^{40}K$ line has been used. Hence, the count spectra have been integrated between 1454 keV and 1461 keV, to take into account the energy resolution evolution between annealings. Finally, the local continuum (quasi flat in this spectral region) has been subtracted. The flux is obtained after dividing by the livetime. The resulting values are displayed in Figure \ref{fig40K} for each revolution, and averaged on $\sim$ 2 months (20 revolutions). We first note that the $^{40}K$ light curve presents a long plateau (between revolutions $\sim$ 600 and $\sim$ 1600) before an unexplained step, that may point out that part of this evolution is not due to the loss of efficiency. We should also consider that the PSD efficiency is slightly varying with the background \citep{roques19}: a decrease of 4\% is observed between revolutions 43 and 885 (corresponding to a first solar minimum), and 4.4\% between revolutions 43 and 1933 (second solar minimum). Thus, while the $^{40}K$ flux evolution reflects an 'apparent' efficiency loss of $\sim$ 30\% at 1460 keV without any supplementary correction, we can derive an upper limit for the detection efficiency loss of $\sim$ 25\% at this energy, after 17 years of operation.

\section{Discussion and Conclusion} 
We have used a  large dataset of observations dedicated to the Crab Nebula
to investigate both the SPI instrument and Crab emission temporal evolutions.
From an instrumental point a view, we have demonstrated that the SPI germanium detector efficiency remains remarkably unchanged (within 5\%) up to 400 keV, after 17 years in space and 33 annealing realizations.  To investigate the efficiency evolution at higher energy, we have used the $^{40}K$ line as an internal source calibration in the 1.5 MeV region. Even though this parameter is less easy to constrain, we have estimated that the efficiency loss is less than $\sim$ 25\%.

Because its calibration matrices rely on an extensive  campaign of ground calibrations coupled to Monte-carlo simulations,  SPI is able to provide an absolute measurement of the Crab  flux and spectral shape above 20 keV, as well as their evolutions over time.
This work completes the studies performed by \cite{crabvar} and \cite{crabComptel} and confirms that
 the Crab emission presents  a modest variability over a couple of year timescale. However,  its steadiness over larger timescales is definitively
  established. The spectral shape also appears to be  stable over time. This outcome deserves to  be examined in the context of the huge flares occasionally observed in the GeV domain. While the   amount of energy dissipated in the Nebula is then important, the SPI observations imply that the  global configuration of the Nebula, as well as the physical mechanisms at work, are not significantly impacted. \\ 
As a concluding remark, we point out that the small deviations  presented by the Crab flux around  the 20 yr mean values  are commensurable with the uncertainties of the instrument calibrations. Consequently, this source remains the best 'standard candle' candidate, in particular for cross-calibration purposes, in the high energy domain. Indeed, it offers an easy way to compare instrument responses, without requiring  simultaneous observations.
 
\begin{deluxetable}{cccccccc}
\tablecaption{Parameters of the histograms of flux and of standart deviations presented in fig. \ref{histo}. \label{val}}
\tablehead{
&\colhead {expected} &\colhead{distribution} &\vline & \colhead {observed}&\colhead{ distribution}\\
\colhead{Period} &\colhead{$F_{mean}$}&\colhead{$<Err>$}&\vline &\colhead{$\mu$} &\colhead{$\sigma$}
}
\startdata
\hline
\hline
&  & 24-50 keV &(units &$10^{-3}$ ph/cm$^2$ s keV)& \\
\hline
 2003-2011  & 6.838  $\pm$ 0.0027   & 0.11  & \vline & 6.826 $\pm$  0.0054 & 0.204 $\pm$ 0.0054 \\
2012-2018 & 6.762 $\pm$  0.0022  & 0.103 &\vline & 6.754 $\pm$ 0.0044   & 0.195 $\pm$ 0.0044  \\ 
\hline
\hline
&   & 50-150 keV & (units  & $10^{-3}$ ph/cm$^2$ s keV)& \\
\hline
 2003-2011  & 1.038  $\pm$ 0.001    & 0.11&  \vline &  1.039 $\pm $  0.0015 & 0.0498 $\pm$ 0.0016 \\
2012-2018 & 1.028 $\pm$ 0.001   & 0.103 &\vline  & 1.028 $\pm$ 0.0016   & 0.0504 $\pm$ 0.0017  \\ 
\hline
\hline
&   & 24-50 keV &(units &standard  deviation)  & \\
\hline
 2003-2011  & 0    & 1  &\vline & -0.35 $\pm$  0.029 & 1.46 $\pm$ 0.029 \\
2012-2018 & 0   & 1 & \vline &-0.32 $\pm$ 0.034   & 1.39 $\pm$ 0.034  \\ 
\enddata
\end{deluxetable}

\begin{deluxetable}{cccccccc}
\tablecaption{Crab nebula best-fit parameters with the Band model for the total dataset and  the two subdatasets (see text and figures). 0.5\% systematic errors were included to data during the fit. Quoted errors are statistical only. \label{fit}}
\tablehead{
\colhead{Period} &\colhead{$\alpha_1$}&\colhead{$E_{c}$}&\colhead{$\alpha_2$} &\colhead{N1}&\colhead{ N$_{ME2}$}&\colhead{F (30-100 keV)}&\colhead{$\chi^2$ (dof)}\\
 \colhead {} &\colhead{ } & \colhead {keV} & \colhead{} & \colhead{ ph /cm$^2$ s keV} & \colhead{/N1 } & \colhead{erg /cm$^2$ s} &\colhead{}
}
\startdata
 1+2~~7.02 Ms  & 1.99  $\pm$  0.01   & 531  $\pm$ 50 & 2.32 $\pm$  0.02 & (7.52 $\pm$ 0.2) 10$^{-4}$& 1.02&   (1.30 $\pm$ 0.01) 10$^{-8}$& 77.5   (47) \\
1~~~~2.73 Ms & 1.99$\pm$ 0.02  & 565 $\pm$ 40 & 2.31 $\pm$ 0.04   & (7.53 $\pm$ 0.2) 10$^{-4}$ & 1.0 & (1.31$\pm$ 0.01)  10$^{-8}$& 60.6  (47)\\ 
2~~~~4.29  Ms       &1.98$\pm$ 0.01  & 507 $\pm$ 40 & 2.33 $\pm$ 0.02   & (7.55 $\pm$ 0.1) 10$^{-4}$ & 1.03  & (1.29 $\pm$ 0.01) 10$^{-8}$ & 79.6 (47)\\
\enddata
\end{deluxetable}

\section{Acknowledgments} The \textit{INTEGRAL} SPI project has been completed under the responsibility and leadership of CNES.  We are grateful to ASI, CEA, CNES, DLR, ESA, INTA, NASA and OSTC for support. The authors thank the referee for her/his constructive comments.

\software{Xspec V12.9.1 \citep{xspec}, SPIDAI (SPI data analysis tool developed at IRAP)  \url{https://sigma-2.cesr.fr/integral/spidai} }


\begin{thebibliography}{}
\bibitem[Arnaud (1996)]{xspec}
Arnaud, K. A. \ 1996, ADASS V, ASP conf. series, Eds G.H Jacoby and J. Barnes, Vol. 101, p.17
\bibitem[Atti\'e et al. (2003)]{attie03}
Atti\'e, D., Cordier, B., Gros, M., et al. \ 2003,  \aap, 411, L71
\bibitem[Abdo et al. (2011)]{flareLAT}
Abdo, A. A., Ackermann, M., Ajello, M., et~al. \ 2011, Science, 331, 739
\bibitem[Band et al. (1993)]{band93}
Band, D., Matteson, J., Ford, L., et~al. \ 1993, \apj,  413, 281


\bibitem[Jourdain \& Roques (2009)]{crab09}
Jourdain, E., \& Roques, J. P. \ 2009,  \apj, 704, 17
 
\bibitem[Kirsch et al. (2005)]{kirsch05}
Kirsch, M. G., Briel, U. G., Burrow, D., et al. \ 2005,  \ Proc. SPIE, 5898, 22
   
\bibitem[Kuiper et al. (2001)]{crabComptel}
Kuiper, L.,  Hermsen, W., Cusumano, G., et al. \ 2001,  \aap, 378, 918

\bibitem[Ling et al. (2003)] {crabBatse}
Ling, J. C. \& Wheaton, W. A., \ 2003, \apj , 598, 334

\bibitem[Madsen et al. (2015)]{CrabNustar2015}
 Madsen, K.K., Reynolds, S., Harrison, F., et al. \ 2015,  \apj, 801, 66
      
\bibitem[Massaro et al. (2000)]{crabMassaro}
Massaro, E., Cusumano, G., Litterio, M.,  and Mineo, T., \ 2000, \aap, 361, 695

\bibitem[Mineo et al. (2006)]{mineo06}
Mineo, T., Ferrigno, C., Foschini, L., et al., .  \ 2006, \aap , 450, 617
   

\bibitem[Roques et al. (2003)]{roques03}
Roques, J. P., Schanne, S., Von Kienlin, A., et~al,\ 2003, \aap 411, L91
\bibitem[Roques \& Jourdain (2019)]{roques19}
Roques, J. P., \& Jourdain, E. \ 2019,  \apj, 870, 92

\bibitem[Strickman et al. (1979)] {crabNRL}
Strickman, M. S., Johnson, W. N. and Kurfess, J. D., \ 1979, \apj , 230, 15

\bibitem[Sturner et al. (2003)]{sturner03}
Sturner, S. J.,  Shrader, C. R., Weidenspointner, G., et~al,\ 2003, \aap, 411, L81 

\bibitem[Vedrenne et al. (2003)]{Vedrenne03}
Vedrenne,  G., Roques, J.P., Schonfelder,  V., et~al,\ 2003, \aap, 411, L63

\bibitem[Weidenspointner  et al. (2003)]{weiden03}
Weidenspointner, G., Kiener, J., Gros,  M.,  et~al,\ 2003, \aap, 411, L113

\bibitem[Wilson-Hodge et al.(2011)] {crabvar}
Wilson-Hodge, C. A., Cherry, M. L.,Case, G. L., et ~al. \ 2011, \apj, 727, L40
\end{thebibliography}
\end{document}